# Hierarchically nanostructured thermoelectric materials: Challenges and opportunities for improved power factors

Neophytos Neophytou[1*], Vassilios Vargiamidis[1], Samuel Foster[1], Patrizio Graziosi[1,2], Laura de Sousa Oliveira[3], Dhritiman Chakraborty[1], Zhen Li[1], Mischa Thesberg[4], Hans Kosina[4], Nick Bennett[5], Giovanni Pennelli[6], and Dario Narducci[7]

[1] School of Engineering, University of Warwick, Coventry, CV4 7AL, UK

[2] Consiglio Nazionale delle Ricerche – Istituto per lo Studio dei Materiali Nanostrutturati, CNR – ISMN, via Gobetti 101, 40129, Bologna, Italy

[3] Department of Chemistry, University of Wyoming, Laramie, 82071, USA

[4] Institute for Microelectronics, Vienna University of Technology, Vienna, Austria

[5] Centre for Advanced Manufacturing, Faculty of Engineering and Information Technology, University of Technology Sydney, NSW 2007, Australia

[6] Department of Information Engineering, University of Pisa, Pisa, I-56122, Italy

[7] University of Milano Bicocca, Dept. Materials Science, via R. Cozzi 55, 20125 Milan (Italy)

[*]N.Neophytou@warwick.ac.uk

# Abstract

The field of thermoelectric materials has undergone a revolutionary transformation over the last couple of decades as a result of the ability to nanostructure and synthesize myriads of materials and their alloys. The *ZT* figure of merit, which quantifies the performance of a thermoelectric material has more than doubled after decades of inactivity, reaching values larger than two, consistently across materials and temperatures. Central to this *ZT* improvement is the drastic reduction in the material thermal conductivity due to the scattering of phonons on the numerous interfaces, boundaries, dislocations, point defects, phases, etc., which are purposely included. In these new generation of nanostructured materials, phonon scattering centers of different sizes and geometrical configurations (atomic, nano- and macro-scale) are formed, which are able to scatter phonons of mean-free-paths across the spectrum. Beyond thermal conductivity reductions, ideas are beginning to emerge on how to use similar hierarchical nanostructuring to achieve power factor improvements. Ways that relax the adverse interdependence of the electrical conductivity and Seebeck coefficient are targeted, which allows power factor improvements. For this, elegant designs are required, that utilize for instance non-uniformities in the underlying nanostructured geometry, non-uniformities in the dopant distribution, or potential barriers that form at boundaries between materials. A few recent reports, both theoretical and experimental, indicate that extremely high power factor values can be achieved, even for the same geometries that also provide ultra-low thermal conductivities. Despite the experimental complications that can arise in having the required control in nanostructure realization, in this colloquium, we aim to demonstrate, mostly theoretically, that it is a very promising path worth exploring. We review the most promising recent developments for nanostructures that target power factor improvements and present a series of design 'ingredients' necessary to reach high power factors. Finally, we emphasize the importance of theory and transport simulations for material optimization, and elaborate on the insight one can obtain from computational tools routinely used in the electronic device communities.

# I. Introduction

Thermoelectric generators (TEGs) are solid-state devices able to convert heat flux arising from temperature gradients directly into useful electrical power. They have the potential to offer a sustainable path for power generation in a variety of industrial sectors from microwatts to tens/hundreds kW, and MW. Their impact could be widespread across many applications including medical, wearable electronics, building monitoring, the internet of things, refrigeration, thermal management, space missions, transportation, and various industrial sectors [1, 2]. TEGs offer significant opportunities to reduce energy usage and $CO_2$ emissions and add towards sustainability, but widespread implementation of the thermoelectric (TE) technology is lacking because of high prices, toxicity, scarcity, and low efficiencies of the prominent TE materials. However, progress on thermoelectric materials has been dramatic over the last several years. Novel concepts and improved understanding of materials growth, processing, and characterization developed during the last few decades have provided new opportunities for the enhancement of the thermoelectric conversion efficiency [2]. Significant progress is not only encountered in the traditional TE materials, but in low cost and abundant materials, which could enable large scale applicability.

The thermoelectric figure of merit $ZT$, which quantifies the ability of a material to convert heat into electricity, has more than doubled compared to traditional values of $ZT \sim 1$, reaching values of $ZT > 2$ in several instances across materials and temperature ranges, and even approaching 3 in some cases [2, 3, 4, 5, 6, 7, 8, 9, 10, 11, 12, 13, 14, 15, 16, 17, 18, 19, 20, 21, 22, 23, 24]. The figure of merit is determined by $ZT = \sigma S^2 T/(\kappa_e + \kappa_l)$, where $\sigma$ is the electrical conductivity, $S$ is the Seebeck coefficient, $T$ is the absolute temperature, and $\kappa_e$ and $\kappa_l$ are the electronic and lattice parts of the thermal conductivity, respectively. The product $\sigma S^2$ in the numerator of $ZT$ is called the power factor (PF). Different materials have optimal $ZT$s at different temperatures, but since the $ZT$ is proportional to temperature, the higher $ZT$ values are commonly observed at the higher temperatures. Efforts are directed to realizing multiple materials with high $ZT$s at different temperatures as application temperatures vary, and not one material can fill the application gap across temperatures.

The recent improvements in *ZT* are mostly attributed to drastic reductions of the lattice thermal conductivity in nanostructured materials and nanocomposites, which has reached amorphous limit values at $\kappa_l$ = 1-2 W/mK and below [2, 4, 25, 26, 27, 28, 29, 30, 31]. The phonon spectrum of most potential TE materials is composed of phonons of a continuum of mean-free-paths (MFPs), spanning from nanometers to micrometers, even up to millimeters in some cases. Nanostructuring targets the dominant phonon MFPs, which differ within materials and operating temperatures, but are typically found to be at the nanoscale. It is certainly the case that the more disorder introduced in a material, the lower its thermal conductivity will be. One of the most successful strategies to reduce the thermal conductivity is to hierarchically nanostructure the materials, such that different size features scatter more effectively different groups of phonon MFPs, as shown in Fig. 1a [32]. In this way a comprehensive reduction of thermal conductivity from phonons in the entire spectrum is achieved [33]. Atomic-scale defects, i.e. defects of up to a few nanometers (mostly atomic defects, but also ultra-scaled quantum dots, second phase islands and even alloying), can effectively scatter short wavelength phonons. Nanoscale defects (e.g., dislocations, alloying, nano-precipitates, larger quantum dots and second-phase islands) can scatter short and medium wavelength (up to ~100 nm) phonons. Micro and mesoscale defects (e.g., grain boundaries) can scatter long wavelength (up to ~ 1 mm) phonons. Grain boundary scattering could also be more effective at higher temperatures [34, 35]. Despite the fact that nanostructures can also suffer from stability and reliability issues, it is currently considered as the most promising way forward. TE materials should be designed to sustain the conditions they are supposed to handle regarding elevated temperatures and temperature gradients, and at some point this needs to be seriously addressed.

While extreme nanostructuring allowed the realization of multiple materials with *ZT* > 2, this achievement was accompanied by reductions in the electronic conductivity, which is also degraded in the presence of defects. In most cases the Seebeck coefficient experiences a slight improvement, but the power factor commonly drops. *ZT* improvements are achieved because the drop in the thermal conductivity is much larger compared to the drop in the electrical conductivity. The difference in the reduction of thermal versus electrical conductivity resides in the differences in the intrinsic MFPs for scattering of phonons versus those of electrons. The dominant MFPs of phonons are typically an order of magnitude larger than those of electrons, thus phonon transport suffers more in the presence of nanostructuring. Some of the best results in terms of *ZT* improvements, however, were achieved in cases where care was taken to avoid power factor reduction [4, 17, 36, 37, 38, 39, 40, 41, 42, 43]. This is commonly achieved in two ways, either by: i) aligning the band edges of the nanoinclusions with those of the pristine material to keep the electronic conductivity high, or ii) going towards the reverse direction, by introducing energy barriers that result in energy filtering and improve the Seebeck coefficient. During the energy filtering process, only the high energy, hot carriers propagate through whereas the cold, low energy carriers are blocked.

Nanostructuring has now allowed materials with ultra-low thermal conductivities, well below the amorphous limits, and very high *ZT*s, but it is gradually running out of steam in providing further reductions to the thermal conductivity. It is becoming increasingly clear that any further benefits to *ZT* must now come from power factor improvements. However, the efforts towards power factor improvements are overshadowed by those towards thermal conductivity reductions. The lack of progress in the power factor is attributed to the adverse interdependence of the electrical conductivity and Seebeck coefficient via the carrier density, which proves very difficult to overcome.

Current research efforts in improving the power factor aim towards identifying materials with favorable bandstructure features, such as resonant states and low-dimensional 'like' features within bulk materials [44, 45, 46], or bandstructure engineering such as band-convergence strategies [45, 46, 47, 48, 49]. Material designs that take advantage of modulation doping and gating were proposed in the past, but in practice never delivered high performance results [50, 51, 52, 53, 54, 55, 56, 57, 58]. Concepts that take advantage of the Soret effect in hybrid porous/electrolyte materials also begin to emerge [59]. In a recent pioneering work, extremely high powers factors were demonstrated in a topological Heusler-based thin layer material [20]. In nanostructured materials, however, the most explored direction is that where the Seebeck coefficient is increased by utilizing energy filtering designs in nanocomposites and superlattices. Design sweet spots can be found for which the power factor is improved, but even so not significantly [60, 61, 62, 63, 64, 65, 66, 67, 68, 69, 70, 71, 72, 73, 74, 75]. All these approaches target improvements either in the Seebeck coefficient or the electrical conductivity, with the hope that the other quantity will not be degraded significantly.

In light of the large progress brought by nanostructuring in TE materials, this colloquium reviews the current progress in power factor improvement efforts and proposes novel design directions worth exploring by experiment. These are directions that are identified through involved simulations in hierarchically nanostructured materials, which constitute the basis of this work, with experimental backing in some cases. In this way, the low thermal conductivities can be combined with high power factors, leading to possibly unprecedented *ZT* values. Specifically, we describe theoretical and experimental findings that can lead to designs in which *simultaneous* improvements in both the electrical conductivity and the Seebeck coefficient are achieved in order to largely improve the power factor $\sigma S^2$. Experimental works have indeed verified that it is possible to achieve very high power factors (PFs > 15 W/mK$^2$, 5× compared to bulk values) in nanostructured Si-based materials [76, 77, 78, 79, 80]. Our simulations show that design optimization can allow for even higher, PFs > 30 W/mK$^2$ [81]. Si-based nanostructured systems have raised significant interest as a platform to test concepts [82]. More recently, however, a surge in efforts to use energy filtering and design the grain/grain-boundary system efficiently in a variety of materials has emerged [73, 74, 75, 83]. Thus, we will provide a recipe on the design 'ingredients' of a generalized potential well/barrier design concept that allows very large PFs. It turns out that optimizing such structures requires thorough understanding of electronic transport in nanostructured media. For this, large scale simulators are needed, and we briefly discuss examples of 'real-space' simulators and models that are suitable for such studies.

The colloquium is organized as follows: Section II describes the main efforts in designing nanostructures for improved Seebeck coefficients, the methods and state of the art. Section III describes the optimal design barrier (or grain boundary) region for effective utilization of the energy filtering mechanism. Sections IV and V describe how PF improvements can be achieved in the presence of hierarchical disorder. For this we discuss the situation where defects are introduced into the grains in addition to the grain boundaries that surround them. Section VI provides a 'recipe' for exceptionally high power factors in an optimal design of the well/barrier (grain/grain boundary) region. Section VII concludes the work and provides thoughts for future directions.

## II. Nanostructuring: materials, methods, and state-of-the-art

The main activity in thermoelectric research in the 1990s was to explore two approaches: i) new materials with low thermal conductivities, and ii) low dimensional systems for improving the PF. The latter is the beginning of nanostructuring, and originated from the prediction by Hicks and Dresselhaus that low dimensional structures could provide higher Seebeck coefficients [84, 85, 86]. In the 2000s, the two approaches began to merge [87]. Low-dimensional features are introduced into materials to form nanostructures, which significantly reduces $\kappa_l$, although the accompanied improvements to the Seebeck coefficient were never realized for reasons discussed in the literature by us and others [88, 89]. The potential barriers at the interface between the nanoinclusions and the matrix material, however, contribute to energy filtering processes, which increase $S$, something that was realized even earlier [90, 91]. In addition to efficiency and $ZT$, the output power is in fact equivalently as important, or even more important when the heat source is unlimited (such as solar heat), or if the heat source is free (waste heat from automobiles and steel industry) [92]. For power generation, when the amount of output power is the desirable outcome, the PF can be more important than efficiency. Furthermore, having higher PF at the same efficiency can provide better thermo-mechanical stability [93].

One of the first successful demonstrations of energy filtering through interfacial engineering was the case of $In_{0.53}Ga_{0.47}As$/InGaAs superlattices which achieved $ZT = 1.5$ [94]. For high $ZTs$ to be reached, other than providing energy filtering, surface roughness engineering and a high density of surfaces was able to decouple the PF and thermal conductivity. In fact, in metal-based InGaAs/InGaAlAs superlattices, computation showed that such superlattices with high barriers can achieve a large $ZT > 5$ at room temperature [60]. Another early example was the "phonon-blocking/electron-transmitting" concept, realized in $Bi_2Te_3$/$Sb_2Te_3$ epitaxial thin films superlattices with thickness structure [10Å/50Å], which reported a record $ZT \sim 2.4$ at room temperature [95], (although not yet reproduced).

Despite the initial bottom-up demonstrations, top-down approaches, such as mechanical alloying and in particular high energy ball milling, are now widely used methods to synthesize nanoparticle powders and bulk-size TE materials. Ball milling is a simple process which directly uses the constituent elements as starting materials, followed by hot pressing sintering (typically current assisted) to melt the powders into nanostructured bulk-size materials. There are several sintering techniques such as spark plasma sintering (SPS), or pulse activated sintering (PAS), or field activated sintering (FAS) [96, 97, 98, 99, 100]. Nanostructuring in this way allowed significant improvements in traditional TE materials mainly through reduction in $\kappa_l$, but also through non-negligible PF increases in a few cases.

Prominent examples are as follows mostly for the traditional Bi/Te/Sb system: In p-type $Bi_{0.4}Sb_{1.6}Te_3$, a 40% $ZT$ enhancement from 1 to 1.4 was observed by applying ball milling and hot pressing to an ingot [8, 101]. The new nanocomposite was characterized by multi-scale phonon scattering centers of fine grains of 50 nm- 2 μm, nanoinclusions of 5 – 20 nm, and atomic defects with sizes less than 5 nm [102]. For the same material, Xie et al. [103, 104] used melting-spinning, hand-milling, and spark-plasma-sintering to synthesize $Bi_{0.52}Sb_{1.48}Te_3$ bulk alloy and obtained $ZT \sim 1.5$ at room temperature. In Ref. [105] $BiSbTe_3$ hierarchical nanostructures were synthesized and nanosheets of thickness around 60 – 70 nm and of length 500 – 600 nm were obtained. The structure also contained spherical nanoparticles of size 150 nm, achieved by tuning the capping agent concentration. Enhanced Seebeck coefficient was observed, due to energy dependent scattering of charge carriers at the nanograin interfaces and energy filtering. In Ref. [106], high interface to volume ratio in $Bi_2Te_3$ was achieved with superassembly-on-epitaxy of $Bi_2Te_3$ nanostructures. This created a material with high density surface and interfaces and epitaxial channels, which allowed partial decoupling of the thermal conductivity and electrical conductivity. $Bi_2Te_3$ super-assemblies of different shapes and sizes were achieved by controlling the deposition temperature and ambient pressure with a power factor significantly increased compared to most nanostructured films. In another example, the TE performance of PbS was largely improved with the incorporation of PbTe inclusions of a few nanometers in size [107]. The thermal conductivity was strongly reduced, whereas the electrical conductivity suffered only a minor cost. In a recent work, $ZT = 1.4$ was achieved in Bi-doped PbTe with the introduction of the $Cu_{1.75}Te$ nanophase, which resulted in reduction in $\kappa_l$ and improvements in the electrical properties of the material [108]. Furthermore, in the case of nanostructured $p$-type SiGe, a $ZT \sim 0.95$ at 800 °C was achieved, 50% larger than bulk [109]. Thus, the ball milling, fast sintering route can effectively create nanocomposite materials (a term that describes a material with multiple and different nano-features). The finalized materials using this method are nanocrystalline, but with the additional incorporation of nanoinclusions embedded in grains and grain boundaries [110].

Ball-milling as well as melt-spinning techniques are also used to achieve nanostructured silicide materials as an alternative low cost TE material for medium to high temperatures [83, 112]. $Mg_2Si$ [113, 114], and metal silicides including $CrSi_2$ [115], $CoSi_2$ [116], $TiSi_2$, $VSi_2$ [117], and $YbSi_2$ [118] have been extensively studied. Nanostructuring methods can produce nanoscale precipitates in the Si matrix, which reduces $\kappa_l$ drastically without affecting the electrical conductivity significantly, and allows for $ZT$ improvements. Importantly, the nano-precipitates are located within larger grain formations, which constitutes a hierarchically nanostructured geometry. The size of the precipitates within the grains can be controlled by adjusting the cooling rate of the melt, and can vary from several nanometers to hundreds of nanometers, whereas the shapes can vary from spherical, to ribbons, to nano-lamellar structures [117].

Other techniques that are used to produce nanocomposite materials are melt-spinning plus spark plasma sintering [103, 119], spark erosion plus spark plasma sintering [120], and chemical metallurgy methods [121, 122]. Review papers that describe all these methods in detail exist, and the reader can refer to those [123]. In one of the most successful examples of nanostructuring by crystal growth and power metallurgy, Biswas *et* al. were able to nanostructure PbTe/SrTe in what they referred to as 'hierarchical' or panoscopic nanostructuring, which included atomic-, nano-, and meso-scale defects to target a range of phonon MFPs and reduce the lattice thermal conductivity across the spectrum [4, 5, 124]. A $\kappa$ of 0.9 $Wm^{-1} K^{-1}$ at 915 K and a record $ZT$ of ~2.2 was demonstrated by the introduction of 1–17 nm SrTe nanoscale precipitates and 0.1-1 μm grains in Na-doped

PbTe matrix. More recently, using this method for the p-type $Pb_{0.98}Na_{0.02}$Te-SrTe system, Tan et al. reported an even lower lattice thermal conductivity ($\kappa$) of 0.5 W K$^{-1}$m$^{-1}$ and a higher ZT of 2.5 at 923K [17]. This multi-scale, or all size phonon scattering center inclusion, is now a widely used approach to reduce $\kappa_l$, although these features are still not arranged in a controllable way. The efficiency in reducing the phonon transport depends on the distance between the nanoinclusions or the second phase segregations, they should be closer than the phonon MFP [4, 43, 125]. Importantly, in these works care was taken such that the insertion of nanoinclusions does not degrade the power factor significantly by arranging for the alignment of the band edges of the constituent material phases such that electronic transport is not noticeably interrupted [4, 5, 124]. This can be done by structural iso-valent substitutions in the lattice using atoms with different atomic weight, like Sr instead of Pb in lead tellurides [4], or Zr and Hf instead of Ti in TiNiSn half-Heusler compounds [37, 126]. The electronic scattering depends on the difference in the energy levels and bandgaps of the parent compounds – the smaller it is, the weaker the electron scattering is. Another successful example of this electronic level matching strategy can be found in Ref. [127], where a careful selection of CdS nanoinclusions in p-type PbSe allowed ZT ~ 1.6 at 923 K. In addition, high average ZT over a wide temperature range was reported in PbSe based systems by tuning the electronic properties by nanostructuring [128]. In $Cu_xPbSe_{0.99}Te_{0.01}$ polycrystalline specimens, a record-high average ZT ~ 1.3 over the whole 400 K to 773 K range, has been reported. Alloying Te to the PbSe lattice and introducing excess Cu to its interstitial voids, resulted in a significant charge transfer from the Cu atoms to the crystal matrix, without degrading the mobility, which consequently improved electrical conductivity.

The initial work on nanostructuring the material for reduction in $\kappa_l$, provides the motivation to also use nanostructuring to design the electronic transport channel with appropriate inclusions for energy filtering and modulation doping [43, 129, 130, 131]. Some examples based on traditional TE materials that take advantage of these effects begin to emerge. In Ref. [132], hierarchical nanostructured $Bi_2Te_3$-based nanowire materials with special interface design were synthesized by sintering into pellets by spark plasma sintering (SPS). The interface states reduce the thermal conductivity, but also influence the electronic transport. A barrier of 71 meV was built at the interfaces, which lead to energy filtering and improved PFs. In another study, $Bi_2Te_3$ and $Bi_2Se_3$ nanoflakes were mixed into sintered pellets, and the band bending at the interfaces created a Schottky barrier which allowed energy filtering with significant *ZT* improvement [133]. However, the relation between the barrier height and filtering needs to be appropriate. A very high barrier will damage electronic transport, whereas a low barrier will offer no filtering advantage and could also allow high thermal conductivity. In other examples, the *β*-$Zn_4Sb_3$ system, embedded with $(Bi_2Te_3)_{0.2}(Sb_2Te_3)_{0.8}$ resulted in a 30% increase in the PF as a result of energy filtering and a *ZT* ~ 1.1 at 648 K was achieved [134, 135]. Similarly, an interface between *β*-$Zn_4Sb_3$ and $Cu_3SbSe_4$ creates an effective barrier produced from band bending which improves the PF [136]. Again in *β*-$Zn_4Sb_3$ with Cd-doping, 10 nm nanosheets (or nanopellets) are produced, which then grow into ~ 1 μm clusters, providing opportunities for filtering [137]. Significant improvement of the PF up to 40% has also been attained in *β*-$Zn_4Sb_3$ [138]. It was demonstrated that the *ZT* of Pb-doped *β*-$Pb_{0.02}Zn_{3.98}Sb_3$-based composites with $Cu_3SbSe_4$ nanoinclusions yields a figure of merit *ZT* = 1.4 at 648 K [138]. This was due to the combination of resonant distortion of the electronic DOS in the Pb-doped matrix and enhanced energy filtering at the heterojunction potential barriers.

The concept of synergistic scattering was also proposed, in which case a semiconductor matrix material dispersed with metallic nanoparticles and ionized impurity dopants provides a synergistic effect for scattering on the potential barriers and the ionized impurities [139, 140]. The carrier scattering becomes strongly energy dependent, which increases the Seebeck coefficient. Another nano-feature that has been widely explored are nanopores, which drastically reduce thermal conductivity [26, 28, 141, 142, 143, 144, 145, 146, 147, 148, 149]. Porosity could lead to an increase in the Seebeck coefficient due to the increase of the entropy per charge carrier due to energy filtering at the pore boundaries, but as we will show later on, this improvement is not significant. It is shown, however, that the presence of hollow pores with multiscale hierarchical disorder leads to more considerable enhancement in the thermopower over its bulk value [79, 150].

Compared to top-down methods like crystal growth from melt or ball milling of ingots, bottom-up methods are much more controllable and versatile in compositions and microstructures [151]. Among the bottom-up methods, the solution-processed methods for nanostructure synthesis have the advantages of low cost, mild preparation conditions, and compatibility with large scale industrial chemical synthesis. It is also a convenient route to tune the nanostructured features to the sub-10 nm range with a precision that cannot be achieved with top-down approaches. Approaches to construct heterostructures involve metal-semiconductor composites, and semiconductor-semiconductor composites. When designing the energy barrier at the interface between heterogeneous interfaces it is crucial to choose an appropriate second phase [151]. Generally, a large barrier is harmful for the electronic conductivity, while it increases the Seebeck coefficient. The choice of materials with appropriate workfunctions can create optimal barriers. For example, solution processed nanostructured chalcogenide $Ag_x$-$Te_y$-$Sb_2Te_3$ TE materials indicated that the optimal barrier is of the order of 100 meV, in which case the PF was improved [152]. However, when designing the barrier interface, it is common to use the bulk material workfunctions to estimate the barrier. This can be inaccurate, as the energy bands of the nanostructured material can differ significantly.

Direct mixing of two semiconductor materials is a general method for synthesis of two semiconductor nanocomposites by directly mixing the solution-processed materials. After stripping of the organic ligands, the nanostructures are pressed into a pellet or deposited as a thin film. This method allows flexibility in controlling the material type and composition [133, 153, 154]. Using this method, power factor improvements have been achieved in the $Bi_2Te_3$-$Bi_2Se_3$ system through energy filtering.

With respect to doping, in general, the techniques that are used to introduce doping into nanostructures are challenging. In chalcogenides, the surfactant-induced doping technique is used to introduce a second component into the matrix TE material by surface-chemistry control, used within the solution processed synthesis [155, 156]. Doping effects can also appear at the heterogeneous interfaces between mixed materials [157, 158]. The Ag-PbS system is an excellent example of how metallic Ag nanodomains contributed to increasing the electron charge of the PbS matrix material by forming potential wells due to the specifics of the band alignment. This was shown to provide improvements in the conductivity and a *ZT* of 1.7 at 850 K [158]. The challenge here, is that the defects concentrate at the interfaces, which limits the doping levels that can be achieved [151]. In other cases, doping is achieved by incorporation of larger species. For example, carbon/graphene doped $Cu_2Se$ materials have recently demonstrated a high *ZT* of ~ 2.4 at 850-1000 K [24, 159, 160, 161, 162].

### III. Nanostructured grain/grain-boundary design for improved PFs

A schematic of a typical nanostructure with hierarchical disorder is illustrated in Fig. 1a. Grain boundaries are common in most TE materials. Nanoinclusions (nano-scale features of up to several nanometers) and atomic defects also scatter phonons, each acting most effectively on different phonon MFPs and wavelengths. Their properties influence transport significantly, as the phonon experiences an interface/discontinuity that causes diffusive reflection or transmission [25, 29, 30, 31, 34, 163, 164, 165, 166]. Typically, grain boundaries and interfaces scatter long-wavelength, low-frequency phonons more effectively (the ones which largely contribute to thermal conductivity) [30], while nanoinclusions, pores and atomic defects scatter short range, high frequency phonons, typically with a rate ~ $\omega^4$, where $\omega$ is the phonon frequency. This is of course largely a function of the size of each feature [167], as mentioned earlier.

Electrons experience a resistance in their transport as well, as these defects essentially introduce discontinuities between different material phases. These form potential barriers/wells observed by electrons. The probability of transmission of an electron over a potential barrier is roughly exponentially reduced with the barrier height, thus a significant resistance is introduced. Setting energy filtering aside, a goal when nanostructuring is to have nanoinclusions for which the band edge discontinuity is as small as possible, or even zero as shown in Fig. 1b, such that the majority electrons (or holes) experience as little resistance to their path as possible. Indeed, some of the highest *ZT* materials take account of this effect, achieving low thermal conductivity without paying penalty on the electronic conductivity [5, 168]. A barrier is beneficial if introduced for minority carriers, because it would contribute in reducing minority carrier transport and bipolar effects at high temperatures [169, 170].

A potential well (negative potential barrier) for the majority carriers can also impede electron flow, though in a minor way [71]. Despite the fact that the charge carriers will have the required energy to flow over the barrier, still, electrons are quantum mechanical objects, and undergo quantum reflections when they experience a change in the potential profile around them, albeit weak in the case of negative barriers. In addition, deep and long quantum wells could trap electrons, as they can emit optical phonons and lose energy, which lowers the current energy and the Seebeck coefficient as well as the electrical conductivity. In general, avoiding band edge discontinuities between the matrix material and the grain boundaries or the different inclusions, allows the reduction of the thermal conductivity with minimal penalty in the electrical conductivity. Potential barriers, however, is the norm in nanostructured TE materials rather than the exception and cannot be easily avoided [171, 172]. This creates a compromise in the design of nanostructures, where the density of defects needs to be large enough to stop phonons of various MFPs, but small enough to allow high electrical conductivity and PF. In most cases where nanostructuring is introduced in a bulk material, however, the electrical conductivity is slightly reduced, the Seebeck coefficient slightly increased, but the overall effect is that the PF is slightly reduced (i.e. the effect on the conductivity is stronger than the effect on the Seebeck) [173].

Energy filtering: The increase in the Seebeck coefficient experienced in the majority of nanostructures is expected, since the Seebeck coefficient commonly follows the inverse trend of the electrical conductivity. It is usually a consequence of energy filtering, under which the low energy, cold carriers are blocked by the potential barriers and are filtered out. What participates in transport are the high energy, hot carriers, that can overpass the potential barriers. The Seebeck coefficient, can be shown from Boltzmann Transport theory (BTE) to be proportional to the average energy of the current flow $\langle E \rangle$ with respect to the Fermi level, $E_F$, as $S = \left[ \langle E \rangle - E_F \right] / q_0 T$. This is a quantity related to entropy and is essentially the ability of a material to separate low energy cold carriers from high energy hot carriers. As indicated in Fig. 1c, electrons with energy above the barrier $V_B$ can overpass the barrier. The majority of mobile electrons, which reside around the Fermi level, can also absorb an optical phonon, gain energy, overpass the barrier and then relax down to the Fermi level by emitting an optical phonon of the dominant energy in the specific material. Tunneling could weaken this process by allowing flow of low energy carriers through the barriers, thus effective filtering barriers need to be more than 3 nm in thickness [69, 70].

Energy filtering is observed in the majority of nanocrystalline materials, and a common example of controlled well/barrier designs to take advantage of energy filtering is the use of superlattices (SL) [60, 61, 69, 174]. Nanocrystalline materials in which carrier flow happens between grains and potential barriers at grain boundaries can also be thought of as 'effective' superlattices at first order. Such nanostructures allow increases in the Seebeck coefficient [174, 175], and interestingly, in some cases cause increase in the PF as well [76, 80, 81, 94, 176, 177]. Initially, however, the primary reason for considering such structures was the reduction of the thermal conductivity as a result of extensive phonon-boundary scattering [178, 179]. At first order, the electronic transport in the superlattice or nanocrystalline material can be seen as transport through a series of independent regions, whose resistivity adds to the total resistivity, and the Seebeck coefficient adds to the total Seebeck coefficient as:

$$\frac{v_{\text{tot}}}{\sigma_{\text{tot}}} = \frac{v_{\text{W}}}{\sigma_{\text{W}}} + \frac{v_{\text{B}}}{\sigma_{\text{B}}} \tag{1a}$$

$$S_{\text{tot}} = \frac{S_{\text{W}} v_{\text{W}} + S_{\text{B}} v_{\text{B}}}{v_{\text{W}} + v_{\text{B}}} \tag{1b}$$

where the weighting factor $v_i$ is the volume of each region. For the nanograins or SL barrier material to have any significant contribution to the Seebeck coefficient, they need to occupy a significant volume of the material, which then causes larger reduction in the electrical conductivity. The combination of a potential well with high conductivity and low Seebeck coefficient, with a potential barrier with low conductivity but high Seebeck coefficient, cannot easily allow for an overall PF which is larger than the largest of the local PFs of the individual regions.

<u>Semi-relaxation of the current energy:</u> On the other hand, improved PFs can be achieved when there is a significant volume of the material in which transport is under non-equilibrium conditions. As electrons travel through the material, in the vicinity of the potential barriers they tend to absorb optical phonons in order to gain energy and overpass the barriers, while after entering the wells again, they find themselves out of equilibrium and emit optical phonons in order to relax back to equilibrium. This process takes place around the barrier within a few energy relaxation mean-free-paths, $\lambda_E$ [72, 180, 181, 182]. In typical semiconductor materials this is several nanometers, which forces electrons out of equilibrium for ~$4\lambda_E$ around the barrier, as shown in Fig. 1d (i.e. the energy of the current flow shown by the blue line is decaying towards its equilibrium position in the well). This region can extend up to 20-30 nm. In that region, the electrons in the wells essentially travel at higher energies than what they would have traveled at in the absence of barriers. Thus, it is as if the high Seebeck coefficient of the barriers is transferred into the highly conductive wells in those regions, which allows for both high conductivity and high Seebeck coefficient. The longer those regions are, the larger the PF improvements will be, in which case materials with weak inelastic processes are favored (but then the Fermi level needs to be risen close to the $V_B$ level). If the barriers are placed as close as possible it helps with reduced relaxation. On the other hand, a large density of barriers will inevitably introduce a large series resistance, thus a compromise is found when the barriers are placed a few $\lambda_E$ apart, in which case semi-relaxation of the carriers' energy is reached. It is from these non-equilibrium regions that significant PF improvements can arise. Indeed, some of the highest power factor values in nanostructured materials with energy filtering were reported in highly doped Si nanocrystalline materials with grain boundaries forming every 30-50 nm, a few times longer compared to the inelastic MFPs [76].

Filtering at degenerate conditions: An important parameter in the optimization of the PF of nanostructures is the position of the Fermi level with respect to the band edge (determined by the doping level). In the case of maximizing the PF of a pristine material, optimal conditions are achieved when the position of the Fermi level ($E_F$) is around the band edge. Inserting barriers in an optimally doped matrix material to begin with (i.e. with the Fermi level at the band edge), almost inevitably reduces the power factor due to reductions in the electronic conductivity. Thus, the Fermi level in the SL or nanocrystalline material needs to be raised depending on the height of the barrier [69, 72, 180, 181], such that the conductivity is regained, and energy filtering is still achieved. The right positioning of $E_F$ allows effective utilization of the energy filtering process, and it can be shown that it allows for PF improvements. The reason is the presence of faster conducting carriers in the wells at degenerate conditions set by the higher $E_F$ level. Figures 2a and 2b show schematics for the energy regions in which charge carriers flow in the cases where $E_F$ is placed: a) at the band edge of a pristine material, and b) at the barrier height of a SL material. In the absence of carrier energy relaxation, electrons flow over the barriers as depicted, but at those energies they have higher velocities. The transmission probability versus energy of electrons in the two cases is computed for an example material using the quantum mechanical non-equilibrium Green's function (NEGF) method including electron-acoustic phonon scattering [71], and shown in Fig. 2c. Clearly, the area under the blue-dashed line, which will determine the current in the SL case, is higher compared to the area under the red line for the current in the pristine material case. In the particular example the PF improvement for the SL structure compared to the pristine structure is ~25% [71]. Thus, the optimal doping conditions for the SL material are not what optimize pristine materials. The Fermi level is elevated, and the doping density needs to be higher to reflect this. Note that here we did not consider energy relaxation of the electrons into the wells to more clearly make the point that filtering at degenerate conditions can provide higher conductivity, but in practice the barriers need to be placed close enough to prevent this.

<u>Filtering from 'clean' low-resistance barriers:</u> The design specifics of the barrier are quite important in achieving PF improvements, and care needs to be taken to mitigate the reduction in the electrical conductivity that they introduce. A key aspect in realizing large power factors when utilizing energy filtering is to reduce the interface and barrier electrical resistance as much as possible. This can be achieved by taking a series of measures, admittedly some of which can be experimentally difficult to control, but nevertheless they can be quite beneficial to the PF and therefore is it worth discussing them.

The first important element in the design of the potential barrier is to leave the barrier region undoped, or 'clean' of dopants and other impurities. Ionized dopants constitute a strong scatterer for charge carriers. In general, at the degenerate conditions that we propose, ionized impurity scattering (IIS) degrades the carrier mobility by at least $5\times$ and needs to be avoided. Thus, the reduction in the conductivity experienced by the introduction of the barriers can be mitigated by making the barrier 'clean' of dopants (we refer to this as the 'clean-filtering' approach [81]). This allows higher mobility carriers at the barrier regions, limited locally by phonon scattering.

Experimentally this can be challenging, however, here we demonstrate the advantage of this design. In Fig. 3 we show the TE coefficients versus carrier density from a simple circuit model in which the electrical conductivity, Seebeck coefficient, and PF are computed as in-series components in a nanocomposite material [81]. The basis structure we consider is a SL with wells of length $L_W$ = 30 nm, barriers of length $L_B$ = 2 nm, and barrier height $V_B$ = 0.15 eV (geometrical features which experimentally showed large power factors in the past [76, 77, 79]). For this we use Boltzmann Transport theory and *p*-type Si parameters such that we match mobility versus density to measured values [76, 77, 81]. The black-dashed line shows the calculation for the pristine p-type bulk Si, with the PF maximum residing around a density of $5\times10^{19}$/cm$^3$. The results shown by the green lines, indicate a structure in which both the well and barrier are doped. Here we consider independent transport in the well and barrier regions (i.e. full energy relaxation in the wells), as in the lower-left inset of Fig. 3b. The barriers in this case cause strong reduction in $\sigma$, a slight increase in $S$ as expected, but finally a strong reduction in the power factor, at least for carrier concentrations up to ~$10^{21}$ cm$^{-3}$. The PF is recovered, however, and even experiences a small improvement if the barrier region is undoped, as a result of mitigated conductivity, as shown by the red lines. On the other hand, the PF is largely improved in the undoped barrier structure if we consider that the energy of the current does not relax in the potential well, a situation depicted in the upper-right inset of Fig. 3b. The results are shown by the blue lines. As explained above, in this case only the carriers with energies above the barrier height propagate, which largely improves the Seebeck coefficient while the conductivity still remains high.

<u>Oblique sidewalls reduce interface resistance:</u> An additional feature of the well/barrier interface which can be proven helpful in mitigating the conductivity reduction is the sharpness of the barrier, or more precisely the lack of sharpness. Quantum mechanically, sharp barriers introduce larger interface resistance as they cause stronger quantum reflections that reduce the transmission probability compared to oblique sidewalls, which have smoother potential profiles. This becomes stronger as the height of the barrier increases. Fortunately, oblique sidewalls are practically more easily achieved compared to sharp barrier sidewalls, as there will always be intermixing between the atomic species of the two phases that form the nanocomposite. In Fig. 4 we show NEGF simulation results for the TE coefficients of different shapes of barriers. The energy of the current flow, which determines the Seebeck coefficient, is indicated by the different lines which correspond to the different barrier shapes in Fig. 4a. Energy relaxation is included in the simulations, which creates the non-uniform energy shape. The conductance, $G$, of these barrier channels is plotted versus the sidewall distance, $d$, in Fig. 4b. At the first instance where the sidewalls become more oblique, as a result of reduced quantum reflections from the barriers, an overall reduction in interface resistance is observed, which improves the conductance by ~20%. For large $d$, the conductance remains almost constant, whereas the Seebeck coefficient increases slightly, because oblique sidewalls effectively shift the conduction band on average higher in energy compared to the Fermi level. Overall, the introduction of the sidewalls increases the power factor monotonically, up to values of ~30% (Fig. 4d) [71].

<u>The validity of thermionic emission:</u> Another aspect of the barrier design is the possibility of thermionic emission when the barriers are thin enough, which would allow in certain cases significantly reduced barrier resistance. This can be inferred from observing the NEGF extracted transmission of the carriers across the barrier as it becomes narrower. This time we include electron-acoustic phonon scattering only (elastic scattering to isolate the effect of carrier relaxation on the barrier from inelastic relaxation processes into the wells). We simulate a channel with a single potential barrier in the middle, and vary the length of the barrier $L_B$ from $L_B = 100$ nm (taking over half the channel) to $L_B = 5$ nm and then to zero, i.e. the pristine channel case (as illustrated in the inset of Fig. 5). In Fig. 5 we plot the energy resolved transmission function ($Tr$) [71, 81]. It can be shown to have a linear dependence in energy at first order in the case of acoustic phonon scattering for a single subband [183, 184]. That linear dependence is captured in the NEGF simulations for the pristine channel (brown line), with the initial point being at the band edge, i.e. 0 eV. It is also captured in the long barrier channel (purple-dashed line), but now the initial point is shifted to the barrier height at $V_B = 0.05$ eV. As $L_B$ is scaled, however, there is a clear shift at energies after $V_B$ towards the $Tr$ of the pristine channel. The 'jump' in the $Tr$ after $V_B$ in the shorter $L_B = 5$ nm barrier channel, clearly indicates that carriers 'see' the barrier, but for energies above the barrier they have a $Tr$ more similar to that of the well. This would be an indication of thermionic emission, in which case the carriers do not 'relax' (at a large degree) on the bands of the barrier, but propagate with the well attributes. Essentially, this says that the carriers with energies above the barrier have transmission probability closer to 1, but the ones with energies below the barrier are completely blocked. Energy filtering in this case is accompanied with less resistance from the barriers.

Thus, to conclude this section, energy filtering is beneficial to the PF when the barriers are clean from impurities, thin enough to promote thermionic emission, and without sharp features. The size of the wells/grains is of the order of several energy relaxation MFPs. The Fermi level is elevated to the level of $V_B$.

## IV. Nanoinclusions, voids and the power factor

Hierarchical nanostructure geometries involve grain/grain boundary geometries, but with the incorporation of additional features, such as nanoinclusions (NIs) of nanometer sizes. While the impact of nanoinclusions on the thermal conductivity is well documented and anticipated [185, 186], their impact on the power factor is just beginning to be understood. Nanoinclusions essentially introduce mostly potential barriers (or wells sometimes) or regular or irregular shapes and sizes, of scattering centers in the material matrix. In most experimental observations, a small improvement in the Seebeck coefficient is observed, along with a larger reduction in the electrical conductivity, such that the power factor is slightly reduced [187, 188, 189, 190, 191, 192, 193]. The effect on the PF, however, is typically smaller compared to the reduction of the thermal conductivity, such that *ZT* typically increases [187, 188, 191, 194].

From a theoretical point of view, the complexity of electronic transport, combining semiclassical effects, quantum effects, ballistic and diffusive regimes, as well including geometry details, makes accurate modelling a difficult task. Models based on the Boltzmann transport formalism and carrier emission over potentials barriers, including band bending and energy dependent scattering appeared in the last few years, and provide first order understanding and adequate match to measured data in some cases [62, 195, 196]. Methods such as the NEGF can provide deeper and more general insight into electronic transport, although they are computationally expensive, especially when describing channels of sizable dimensions (both in length and width). We have performed such NEGF simulations in geometries with NIs [183]. In the inset of Figure 6c we show a schematic of a regular array of NIs of 3 nm in diameter, which is a typical size found in experiments. We simulate a dense, hexagonally oriented network, in order to maximize its effect on transport properties. Whether the network is regular or randomized, makes only little difference to the PF [197].

Potential barriers formed by NIs, similar to those formed in SLs or by grain boundaries, could be expected to also provide energy filtering and improve the Seebeck coefficient, and hopefully the PF [198, 199]. In the design space, apart from the geometrical configuration of the NIs, other important parameters are the barrier height that the NI introduces in the conduction/valence band of the matrix material, and the position of the Fermi level. These are the parameters that have strong influence on the electrical conductivity. Figures 6a-c show the simulations for the thermoelectric coefficients $G$, $S$ and $PF$ ($GS^2$), versus the NI barrier height $V_B$ for various positions of the Fermi level $E_F$. For each Fermi level, we vary the NI barrier height from $V_B = 0$ eV to $V_B = 0.2$ eV. These are similar band offsets that one encounters in promising TE materials, for example, PbSe/CdSe with a valence band offset of 0.06 eV, PbSe/ZnSe and PbS/CdS with a valence band offset of 0.13 eV [5]. The conductance $G$ in Fig. 6a shows the expected decrease for all Fermi levels as $V_B$ is increased due to the potential barriers blocking the electron flow, but a saturation for higher $V_B$. Similarly, the Seebeck coefficient in Fig. 6b increases with $V_B$ at the order of 20%, before it saturates as well. NIs offer weaker energy filtering compared to SL barriers.

It is interesting that even in the case of such a dense NI network, neither $G$ nor $S$ are affected significantly. These simulations align quantitatively with the majority of experimental observations as well. The corresponding power factors are shown in Fig. 6c. Clearly, the highest PF is observed in the case where no NIs are placed (i.e. $V_B = 0$ eV) and the Fermi level is aligned with the band edge ($E_F = 0$ eV, green line) as in pristine materials, which also seems to agree well with measurements. It is only when the pristine material has a non-optimal Fermi level to begin with, that the NIs can introduce a small energy filtering effect, which improves the PF for moderate barrier heights (see red/blue lines). For large barrier heights the PF is even reduced at levels below the corresponding pristine material case for all Fermi levels. Thus, NIs offer only weak energy filtering at best, to improve the PF. In fact, if it was not for the reduction of the thermal conductivity that the NIs cause, NIs should not be used for ZT improvements. This is of course if one considers a material with an optimized Fermi level position at $E_F \sim E_C$ to begin with, which is rarely the case in practice. If one considers, however, that the position of the Fermi level is in general not at the optimal point, then there is a possibility of moderate power factor improvements of the order of ~10% (red, blue lines). The power factor lines in Fig. 6c for $E_F > E_C$ indicate that a PF maximum is reached when $V_B$ is approximately at $E_F$. Raising $V_B$ even further takes away this increase and forces the power factor to saturate at a lower level (to around 50% of the initial PF). Experimental observations with NIs embedded within matrix TE materials also point to the direction of small band offsets to retain high conductivity and PF benefits [134, 136, 187, 200]. In most experiments in which a matrix material is enriched with NIs, the PF is degraded slightly, due to a reduction in the electrical conductivity that improvements in Seebeck are not sufficient enough to compensate.

In terms of the nanostructuring density, the norm is that the larger the amount of nanostrucuturing is, the higher the reduction in the thermal conductiviy. With regards to the PF, the situation is somewhat different. Figure 7 shows the TE coefficients, $G$, $S$, and PF for four nanostructured geometries with different density of nanoinclusions as indicated in the inset. These four simulated geometries consist of: a 2×4 array (green lines), a 4×4 array (black lines), a 6×4 array (blue lines), and an 8×4 array (red lines). The Fermi level is placed at $E_F = 0.05$ eV (dashed-red line in Fig. 7a) and the barrier height $V_B$ is varied. Fig. 7a shows that the conductance $G$ falls as $V_B$ increases, and as the number of NIs in the channel is increased. Likewise, as the number of NIs increases, the small effect of energy filtering they introduce is increased and an improvement in $S$ is observed. The increase is of the order of 10% for the 2×4 channel, and is increased to approximately 25% for the 8×4 channel as seen in Fig. 7b. The power factor in Fig. 7c increases slightly until $V_B \sim E_F$, but as $V_B$ increases even further, the power factor falls to values below the pristine channel value for all channels. Interestingly, for small barrier heights of $V_B < E_F$ the density of NIs has little effect on the power factor, which increases independently of NI density. In fact, a sweet spot can be identified in the PF, as shown in Fig. 7c, at $V_B \sim 0.07$ eV, or $V_B \sim E_F + k_B T$ at which point the PF is completely independent of the NI density. This observation indicates that the density of nanostructured materials with NIs can be optimized for maximal reduction in the thermal conductivity, at little or no cost to the power factor. At higher $V_B$, on the other hand, the detrimental effect of NI density is more important, resulting in a decrease compared to the pristine material power factor. Thus, in the design of materials with NIs, the choice of $E_F$ and $V_B$ are interlinked, with the Fermi level, $E_F$, needed to be placed at the vicinity of the barrier height $V_B$, or somewhat lower. Overall, the NIs do not seem to offer significant opportunities for PF improvements compared to their pristine materials, but on the other hand, they do not degrade the PF drastically either, which is also the case in the majority of experimental works [188, 191, 192, 201].

To sum up, superlattices, or potentials barriers on the boundaries of grains, offer slightly higher possibilities for PF improvements as discussed above. In their plain SL barrier form without additional optimization, improvements can reach up to ~20-30%. It is easier for electrons to effectively find their way around the NIs and avoid them, thus lower PF improvements are achieved in NI structures. In both cases, however, the benefits are moderate, and $ZT$ improvements come from $\kappa_l$ reduction. Later on we will show how other design ingredients can be combined in such geometries to provide large PFs.

As a side note, in most simulation works, the structures considered have a NI geometry and diameter that are set in a periodic way, i.e. regular hexagonal arrays of fixed diameter. In reality the nanostructuring in nanocomposite materials takes random forms. The specific location of the grain boundaries, the NIs, their size, the barrier height, their density, and even the position of the Fermi level cannot be controlled precisely. However, in SLs, variations in the lengths of the well/barrier regions do not affect the power factor significantly. What is detrimental are variations in the barrier heights (that degrade the conductivity) and extremely thin, easy to tunnel barriers (which degrade the Seebeck coefficient) [69, 181]. In the case of NIs, in a similar manner the variability in the geometry and positions of the NIs does not affect the power factor in a noticeable manner [197], but also for low $V_B$ the density and $V_B$ itself do not affect it either.

# V. Hierarchical nanostructuring architectures and the power factor

I. Hierarchical geometry materials

The next step towards hierarchical designs, is the incorporation of both, elongated superlattice/grain boundary potential barriers, in addition to nanoinclusions, or voids in the intermediate region between the SL barriers. These structures can combine the features of SLs and NIs and can be designed to provide: i) benefits to the PF through energy filtering, and ii) immunity of the PF to the presence and density of the NIs. Schematics of the potential barriers in such structures are indicated in the insets of Fig. 8, where the nanostructuring features are represented with the raise in the potential of the matrix material. Such approaches are extremely effective in reducing the thermal conductivity [4, 63, 187, 188, 189, 190, 191, 192, 193, 202, 203], but here we examine the electronic transport behavior of such geometries.

For power factor optimization, the Fermi level needs to be brought high up close to the SL potential barriers, and the potential barriers are arranged close enough, 30-50 nm to allow reduced energy relaxation. Interestingly, at elevated $E_F$ conditions, in a SL structure, the transmission probability of electrons is quite robust to the nanostructuring that is placed in between the barriers, in such a degree to allow the conductance to remain high close to the SL level. Figures 8a and 8b, show the NEGF extracted electronic transmission function versus energy in a SL structure for cases of different NI barrier height, $V_N$, and different NI densities as indicated in the insets, respectively. In Fig. 8a, by changing the NI barrier height, $V_N$, from zero up to 0.1 eV, the transmission changes only slightly. The same is observed in Fig. 8b when the number of NIs changes from zero to 4 and then to 10. These minimal changes to the transmission indicate that the PF performance will be even more robust to the NI density and their details (as the Seebeck would slightly increase). Their density, however, can drastically affect the thermal conductivity by increasing phonon scattering as shown in several works [204, 205, 206, 207, 208]. For example, in Si the MFP for electrons is of the order of few nanometers, but for phonons the dominant MFPs are ∼135 nm - 300 nm [209, 210, 211]. This difference in MFPs largely increases the influence of closely packed NIs on phonons, rather than electrons.

Figure 9 shows the TE coefficients in this hierarchical nanostructuring case (a SL with NIs), as functions of the NI barrier height $V_N$ and for increasing NI number density [71]. We show results for simulations that consider only electron scattering with elastic acoustic phonons, which is the case for the most promising performance. The structure considered contains SL barriers of height $V_B = 0.05$ eV, and we plot data versus the heights of the NI barriers, $V_N$. Each sub-figure shows results for three structures, containing 4, 6 and 10 NIs in the regions between the SL barriers (as shown in the inset of Fig. 9b). In all cases, as the NI barrier height increases, the conductance is reduced, however only weakly. The Seebeck coefficient demonstrates only a small increase, as the NIs tend to be weak scatterers for low energy electrons and provide a small filtering effect. Overall, the PF exhibits a slight degradation of the order of 10% when NIs are introduced, most noticeably when the $V_N$ increases beyond the $V_B$ (Fig. 9c). However, importantly, even at the high NI density and high $V_N$, the PF is higher than that of the pristine material (horizontal dashed-green line in Fig. 9c).

These nanostructuring design approaches could help open the path to the optimization of new generation nanostructured thermoelectric materials by not only targeting reductions in thermal conductivity, but simultaneous improvements in the power factor as well. In that case, benefits to the PF by > 20% can be achieved. An important design ingredient here is the elevated Fermi level close to $V_B$. Elevated Fermi levels not only allow for high conductivity, but they contribute to the immunity of the transmission to the NI density. The conductivity in structures with lower Fermi levels, is not as immune to the NI density, and suffers significantly [212].

Note that when we consider inelastic optical phonon scattering the performance is degraded, even to values lower than the optimal ones for the pristine matrix material we began with. This is because the length of the channel employed is 100 nm, significantly larger compared to the inelastic mean-free-path for electrons, which is set to ~13 nm, and larger than optimal (~30 nm). Thus, a significant degree of energy relaxation can appear in the wells, in which case the Seebeck coefficient is degraded significantly. However, the insensitivity of the PF to the NI density and barrier height is still retained.

<u>The effect of voids:</u> The far right points connected by the black-dotted lines in the sub-figures of Fig. 9, indicate the corresponding results in the case where the NIs are replaced with voids. For simulation purposes, we increase $V_N$ in those geometries to very large numbers, effectively leading to vanishing wave function in those regions, which resembles a void structure. Voids cause significant degradation in the conductance and in the PF; namely, there is 30% - 50% reduction from the SL reference depending on the void number (it turns out that in this case the density has a sizeable effect). The Seebeck coefficient, surprisingly also seems to be reduced slightly in the presence of voids. Interestingly it can be reduced to values slightly below the Seebeck value of the SL channel without NIs or voids that we began with (left-most data points in Fig. 9b). It is important to mention, however, that voids degrade the thermal conductivity drastically, compared to NIs [26, 202, 213]. Thus, despite the ∼50% reduction in the PF, the thermal conductivity is reduced by more than an order of magnitude, with enormous benefits to *ZT* [212]. In fact, nanoporous structures that even take advantage of phononic effects by regular arrangement of pores, could provide drastic reduction of thermal conductivity [214, 215], although coherent effects are still under investigation.

Note that as of now we elaborated on the possibility of forming potential barriers. However, in the myriad of alloys that are examined for new generation TEs, there are possibilities of potential wells forming NIs and SL barriers which can also reduce $\kappa_l$. In that case the power factor is quite resilient under either elastic or inelastic scattering processes. Potential wells cause some obstruction to transport due to reflections at the interfaces of the SL and NI boundaries, but this is not enough to cause significant reduction of the PF. What is required here, however, is that the wells are narrower compared to the inelastic mean-free-path, such that the charge does not relax into them. However, in this case no power factor improvement strategies are possible. Therefore, in materials in which transport is dominated by elastic scattering, or if the inelastic scattering energy relaxation length is similar or larger than the characteristic geometrical features of the channel, it is beneficial to utilize nanostructures that form potential barriers, while setting high Fermi energies at the level of the SL barriers. In the case where the dominant scattering mechanisms are inelastic in the underlying geometry, then nanostructuring using potential wells is more beneficial. Although in this case improvements cannot be achieved, at least the reduction to the PF is insignificant.

## II. Dopant non-uniformity combined with energy filtering

Despite the huge benefits in reducing $\kappa_l$, hierarchical nanostructuring can only provide moderate PF improvements of the order of ~40% at most. However, once this is combined with dopant non-uniformity (or modulation doping), the benefits can be tremendous as we show below. The PF of thermoelectric materials is optimized at highly degenerate carrier densities between $10^{19}$-$10^{20}$/cm$^3$. To achieve such carrier concentrations, equally high level of doping is required, which introduces a strong scattering mechanism and reduces the carrier mobility significantly. Figure 10 shows in red lines the calculated phonon-limited mobility versus density for *p*-type Si (dashed) and *p*-type HfCoSb (solid), a half-Heusler material with promising TE properties. In blue lines we show the calculated mobility under phonon plus ionized impurity scattering conditions, which is much lower for both materials. At the density where the PF peaks, shown by the arrow, the phonon-limited mobility is ~5× higher compared to the IIS-limited mobility. Modulation doping techniques attempt to concentrate the dopant atoms in islands within the material [50, 58, 216]. The mobile charge carriers then flow around the islands with mobility closer to the phonon-limited, rather that the IIS-limited value. This is similar to what is implemented in high electron mobility transistors (HEMTs), where a delta-doping layer is placed in the gate insulator, rather than the channel [217, 218]. Efforts to realize phonon-limited conductivity have also been directed towards gated materials [55, 219, 220, 221, 222, 223, 224, 225]. Similarly to electronic devices, a gate is built on a thin layer or nanowire TE material, and an applied bias creates accumulation of carriers into the channel, which is undoped to begin with. Although such techniques are indeed promising to improve the PF significantly, in practice the theoretically predicted improvements have not been realized. In most cases it is also challenging to realize the proper nanostructured geometry that allows dopant segregation from the conducting channel.

Over the last few years, reports of exceptionally large power factors (~15 W/mK$^2$) have been reported in bottom-up deposited highly-doped p-type nanocrystalline Si material after a special temperature annealing treatment [76, 78]. Further work on a hierarchical nanostructured extension of the polycrystalline material with pores within the grain domains measured an even higher PF (~22 W/mK$^2$) as a consequence of a further improvement in the Seebeck coefficient [79]. Annealing of highly doped Si nanocrystalline materials were also reported in systems with dislocation defects to allow for PF improvements up to 70% compared to the pristine material [80]. A theoretical model based on a combination of energy filtering by the barriers formed at the grain boundaries [226, 227, 228] and dopant segregation was used to explain the behavior [76], and such models also predict that this design direction can provide even higher PFs [81]. These are clear demonstrations that PFs can provide the next leap forward for TE materials.

Based on these observations, a design concept is devised, which combines energy filtering and dopant non-uniformity. Dopant-free regions are assumed to exist in the material, which in this case reside around the grain boundary (potential barrier) but also extend into the potential well. Thus, the potential well is not uniformly doped, but the doping density is higher in the middle of the well, and is reduced (or eliminated) in the regions around the barriers. Below we develop this concept and present its potential as a novel direction of research. Despite its implementation challenges, we show that it can create structures with unprecedentedly high PFs.

The dopant non-uniformity introduces electrostatically a potential barrier at the grain boundaries (in addition to a barrier due to the grain boundary region itself), allowing energy filtering. Practically, as in the structures described above, this could be realized in heavily doped nanocrystalline materials of light dopant atoms. After high temperature annealing, the dopants tend to segregate towards the grain boundaries, where they precipitate, and/or become deactivated [76, 78]. In a more controllable fashion, a possible practical realization is to fabricate 2D superlattices formed of n++/i, or n++/n- junctions [81]. In that case the barriers are formed in the intrinsic or lightly doped regions, which will be regions 'cleaner' of dopants, experiencing phonon-limited mobility. Lithography can be used for the definition of windows through an oxide layer (grown by thermal oxidation, for example) on an SOI wafer, and shaped by lithography and etching to act as a mask for the doping process. Oxide windows, and hence the final doping concentration, can be arranged to form a SL array of dopant variation regions using ion implantation (for example). One can go even further by lithographically defining lines in the x- and y-directions to form a square 'net', and then dopant diffusion can create highly doped islands in the regions between the 'square net'.

The generalized geometry concept is shown in Fig. 11a. It consists of a three-region nanostructured material, in which the barriers are depicted by the grey colored regions and the blue regions represent the heavily doped regions. The red colored regions in between the heavily doped regions and the barrier regions are part of the wells (as in a nanocomposite material, for example), but they are undoped, or lightly doped depending on the process. A simplified schematic of the potential profile in a 1D cross section of the material is shown in Fig. 11b. The oblique potential in the middle region can be a result of the n++/i junction that is formed, pushing most of the depletion region in the undoped, intrinsic part. Without focus on the details of the barrier formation and shape, charge carriers in the undoped well region will primarily have phonon-limited mobility, and the overall structure will have improved conductivity.

In the simplest way, the TE coefficients in the case of the three-region structure can be computed by combining the individual coefficients of the three regions (well-W, intrinsic-i, barrier-B) in-series. The predictions for the TE coefficients of such structures are extremely encouraging. We note here that a grain/grain boundary architecture provides an additional benefit to the Seebeck coefficient, helped by the large ratio of the grain to grain boundary thermal conductivity. Essentially the overall Seebeck coefficient, from heat continuity reasoning, is the weighted average of the Seebeck coefficients in the different regions with the weighting factor being the inverse of the thermal conductivity (which is smaller in the grain boundaries). In Figs. 11c-e, assuming a well of size $L_W = 30$ nm and a barrier of $L_B = 2$ nm, a $W = 10$ nm undoped region around the heavily doped well center, a ratio for the thermal conductivities of $\kappa_W/\kappa_B = 10$, and a barrier height $V_B = 0.15$ eV, the PF reaches the very high value of 17 W/mK$^2$ (red lines). Here we assumed optimal conditions, for which the energy of the electrons does not relax in the well. Even better, if thermionic emission over the barrier is assumed (as in the case of thin barriers), and the different parameters $V_B$, $W$, and $\kappa_W/\kappa_B$ are optimized to push to high, but realistic values, astonishingly high PF values of 30-50 mW/mK$^2$ can be achieved [81]. This is remarkable, also by taking into consideration that such structures will also provide ultra-low thermal conductivities, and can be further enhanced by inserting NIs within the grains (second-phase dots or pores). The TE coefficients of the pristine bulk material is depicted by the black dashed lines, indicating the drastic improvement that this design provides simultaneously to both $\sigma$ and $S$ and finally the PF.

It is interesting to notice the carrier mobility in this design in Fig. 11f. The black-dashed line shows the mobility of the pristine material versus carrier density (or doping). The red-dashed line shows the phonon-limited mobility of the pristine channel. The red-solid line shows the mobility of the nanocrystalline material. At low carrier densities the mobility is low, as is the usual case in nanocrystalline materials. At high doping densities, however, the mobility approaches the phonon-limited value, rather than the phonon plus IIS value, which allows for very high conductivity and PF.

<u>The validity of reduced energy relaxation:</u> In the design proposed, an important aspect is the placement of the barriers in short enough distances from each other in order to prohibit energy relaxation, but as far as possible from each other to prohibit resistance increase from the multiple interfaces. It turns out that at the optimal barrier separation, the energy of the current suffers a degree of relaxation into the wells. It is interesting to examine, however, under what geometrical conditions and under what inelastic scattering conditions the reduced relaxation is satisfied. Figure 12a shows the energy of the current flow extracted from a series of NEGF simulations in a channel geometry of well size $d = 50$ nm, in which only electron-optical phonon scattering is considered, with a strength which results in a MFP of ~15 nm. The different lines show simulation results for different phonon energies, as indicated in the legend, whereas the horizontal lines in the second well indicate the energy at which the energy should fully relax in the pristine case. Obviously the overall energy is higher compared to its pristine level, which signals the improvement in the Seebeck coefficient not only by the barriers themselves alone, but also in the well regions. The Seebeck coefficients of this channel are shown in Fig. 12b by the red line. The blue line shows the Seebeck coefficient in the case where we vary the scattering strength, and both are plotted versus their corresponding MFP. The horizontal black dashed lines indicate the Seebeck coefficients for the pristine well material and the pristine barrier material independently. Clearly, in almost all phonon energy and scattering strength cases the Seebeck coefficient of the nanocomposite structure is at least in the middle region between that of the well and barrier materials, and in many cases closer to the barrier one.

It is important to stress here that it is the regions in which the energy has not yet relaxed that finally bring benefits to the PF. To illustrate this, in Fig. 12c we show the Seebeck coefficient for the material with optical phonon energy 0.06 eV by the red line, but now the well size is changed from 10 nm to 100 nm. Again, even up to 80 nm well size, the Seebeck coefficient is closer to the $S_B$, rather than the $S_W$. These indicate that once correctly designed with sub-100 nm well sizes, nanocomposites can provide much larger Seebeck coefficients compared to what is expected by the weighted average of the volume fractions of the two regions. This simplistic consideration is shown be the magenta line, and predicts much lower overall $S$.

Model for the Seebeck coefficient with energy relaxation: Thus, the simple picture of combining the Seebeck coefficients of the two components scaled by geometrical considerations is not accurate at the nanoscale, and needs to be corrected. A simple model which can provide more accurate estimates can be a useful fast guide to experiments with nanocomposite materials. The model makes use of the energy relaxation length, $\lambda_E$, which can be estimated by fitting the exponential drop of the energy of the current after the potential barrier (see the inset of Fig. 12b). The combined Seebeck coefficient of the overall system is given by:

$$S^{sys} = \frac{\dfrac{L_B^{total} S_B^{bulk}}{\kappa_B} + \dfrac{L_W^{total} S_{W\text{-rel.}}}{\kappa_W}}{\dfrac{L_B^{total}}{\kappa_B} + \dfrac{L_W^{total}}{\kappa_W}} \qquad (2a)$$

where the $S_W$ is replaced by $S_{W\text{-rel.}}$, a quantity which includes the relaxation physics and is given by [182]:

$$S_{W\text{-relax.}} = S_W^{bulk} + \left(S_B^{bulk} - S_W^{bulk}\right)\left(\frac{2\lambda_E}{d}\right)\left[1 - e^{-d/\lambda_E}\left(1 + \frac{d}{2\lambda_E}\right)\right] \qquad (2b)$$

Note that the overall Seebeck coefficient in Eq. 2a is averaged between the Seebeck coefficients of the two materials by the length of each region (which could be replaced with the area or volume depending on the material geometry specifics) and the inverse of their thermal conductivity. The Seebeck coefficient in the well in this case is not uniform, but accounts for higher current energy in the regions adjacent to the barriers before they reach relaxation. Equation 2b, gives the overall Seebeck in the well region of length *d*, by integrating locally the average energy of the current flow. The model is simple as it involves the Seebeck values of the pristine barrier and well materials, as if they were infinite and not connected to each other, the well thickness, *d*, and the inelastic electron relaxation MFP, for example, the electron-optical phonon scattering MFP. In the limits of $\lambda_E \to 0$ and $\lambda_E \gg d$, one can observe that the Seebeck coefficient approaches $S_W^{bulk}$ and $S_B^{bulk}$, respectively, as expected. The presence of relaxation length in the wells indicates that the Seebeck of the barriers extends partially into the wells up to a few inelastic MFPs of energy relaxation.

In macroscale materials, the energy relaxation MFP is insignificant compared to the grain size and does not affect the overall Seebeck coefficient. In that case, the simple geometric averaging of the Seebeck coefficients of the two regions with possibly a slight adjustment could be adequate. In nanostructured materials, however, where the grain size is of the order of the MFP, that region has a major influence and needs to be included in the model, as it is that region which provides most PF benefits. High Seebeck is achieved because of the higher energy of current, and high conductivity because electrons travel at higher energy, more conductive states. In macroscale systems where the inelastic MFP is much smaller compared to the well/grain size and the non-equilibrium region is small compared to the well size, significant PF benefits from energy filtering are not usually observed.

# VI. Simulation tools for electronic transport in nanostructures

<u>Simulation essentials:</u> The design of nanostructured thermoelectric materials requires at this point advanced simulation tools to drive material identification and optimization. The geometry optimization and the specific engineering of the well/barrier interfaces, the doping level, etc. are all interlinked with internal properties of the pristine material, such as the electron-optical phonon relaxation MFP. Because of this, the large number of potential materials and their alloys with different internal properties, are optimized in different ways, i.e. at different nanostructured geometries, doping and barrier specifics. Simplified models for the Seebeck coefficient and the conductivity using circuit analysis could be a first order guess. However, for more complex geometries and designs, simulators that take into account the 'real-space' geometry description are more accurate and predictive [229]. The design 'recipes' described above have either emerged or some of their 'ingredients' have been validated using advanced simulation tools, which can capture the major electronic transport features in the material.

Recently, 'real-space' simulators based on quantum and semi-classical models to address transport in TE nanostructures have been developed. Two examples of such methods are the quantum mechanical NEGF [230] and the semi-classical Monte Carlo [231]. Both methods are borrowed from the electronic device communities, which have developed them and brought them to maturity [232, 233, 234]. Simulators based on these methods can scale to adequate channel sizes or hundreds of nanometers with enough computational power. In their simplest form, which will allow such large sizes, the pristine material is described by an effective mass band, in which electrons undergo deformation potential scattering, requiring a few parameters such as acoustic and optical phonon deformation potentials and optical phonon energies. On the other hand, the TE materials are primarily complex bandstructure materials with multiple, highly warped bands, and complex phonon dynamics. The simplest way forward is to extract appropriate conductivity and density of states effective masses (simple codes that can do that exist [235]), and an approximate value for the optical phonon energies from DFT extracted phonon spectra. Deformation potentials are not yet accessible for most TE materials and very few studies attempt to extract them [236]. At the moment, the simplest way to estimate them is to attempt to fit the measured mobility data to deformation potential theory, which however could include a large error. An accurate methodology to estimate deformation potentials could allow more reliable use of 'real-space' transport codes for hierarchically nanostructured TEs. Finally, we stress that such simulators need to be easily accessible and usable, in the same way that codes which compute the TE coefficients in bulk materials are, such as BoltzTrap or BoltzWann [237, 238].

The NEGF method is being used in several occasions in the literature [230], as it offers a way to describe electronic transport in arbitrary nanostructured geometries. This method treats electrons quantum mechanically, and in the majority of studies it is used in its ballistic transport form for both electrons and phonons. In this case one obtains the quantum mechanical transmission function, and using Landauer's formalism, the TE coefficients are extracted. Ballistic, coherent transport, however, is problematic for transport in nanostructures, as it overestimates the degrading effect of the nano-features in the conductivity. In fact, a disordered channel material using coherent transport will reach localization as the channel length is increased. The way to avoid this is to implement electron-phonon interactions in NEGF, although this can be computationally intensive and requires self-consistent loops, especially if 2D channels (channels of finite width) are simulated. Figure 13 shows the quantum mechanical transmission function from NEGF for the channel with NIs as shown in the inset. The blue lines are for the pristine material case without the NI barriers – the highest staircase transmission is extracted under ballistic coherent conditions. The black lines are for the material with NIs – the highest of the two, which includes spikes, is for the coherent transport case. In the incoherent cases, where the electron-phonon interaction is included, the transmission is smoothed and takes a linear shape, as expected from semiclassical transport simulations as well [184]. Such simulation works and simulators are now beginning to emerge in order to capture the scattering complexities in TE materials [69, 71, 72, 173, 183, 239]. In general, 1D channels are used, however, 2D treatment is essential to treat hierarchical nanostructured geometries, or more precisely channels of some finite width. 3D treatment is computationally prohibitive, and any such studies are limited to extremely narrow channels [240, 241].

Semi-classical Monte Carlo simulations are also starting to emerge in TE materials [242, 243, 244, 245, 246, 247, 248, 249]. The advantage of MC is that the computational cost increases linearly with the system size and scattering events can be incorporated relatively easily. Electrons are initialized in the simulation according to their distribution, and then let to flow in the channel under the influence of forces and scattering events [229, 231, 250]. Monte Carlo, however, can suffer from conversion issues, and as it is a statistical method it can also accumulate noise in the results. In fact, thermal simulations, in which a temperature difference, rather than a voltage difference is applied across the two ends of the simulation domain in order to extract the Seebeck coefficient, seem to be much noisier. It is more convenient to track the energy of the current flow by the electrons in the channel, rather than perform thermal simulations sometimes.

However, both the NEGF and the MC methods, despite some disadvantages and high computational costs, can be important in providing reliable understanding and optimization routes for the PF of hierarchically nanostructured TE materials. We also note here that both methods are also used for phonon transport in TE materials, in which case the *ZT* figure of merit can be extracted. The phonon MC is well established by several groups [213, 251, 252, 253, 254, 255, 256, 257, 258, 259, 260, 261], as well as the ballistic phonon NEGF [262, 263, 264, 265, 266]. The phonon NEGF with anharmonic effects, however, although developed by some groups but primarily for ultra-narrow channels [267, 268, 269], has not yet been expanded to treat large nanostructured TE materials. To-date, the most important method to compute the thermal conductivity in nanostructures remains molecular dynamics, which has been used extensively (mostly for Si/Ge materials) [270, 271, 272, 273, 274, 275, 276, 277, 278, 279, 208, 281, 282, 283, 284, 285]. Molecular dynamics, on the other hand, require accurate interatomic potentials which are only available for a limited range of materials.

Large scale simulators can also be used in developing simple models, that better describe the behavior of nanostructured materials compared to the traditional models. In general, simplified models for scattering on the boundaries of NIs and grains are not reliable enough to reach convincing justifications for the PF trends. Estimating the electronic conductivity and Seebeck coefficient in such systems cannot easily be mapped to experiments, as the details of boundaries are too complex to model accurately. On the other hand, simple models are of course very useful and should be used for easiness, and their validation from more advanced simulations, or through experimental data can get us long way into understanding measurements [62, 76, 182]. In general, the interfaces between materials is an important aspect of the PF optimization, and needs to be understood better.

# VII. Conclusions, outlook and prospective

This colloquium attempts to make the point that hierarchical nanostructuring is a very promising research direction for achieving extraordinary thermoelectric power factors, even more than an order of magnitude compared to the pristine material's corresponding values. This could lead to a leap forward in the field of thermoelectric materials. Hierarchical nanostructuring is an approach currently used to scatter phonons with mean-free-paths across the spectrum to reach drastic reductions in the thermal conductivity, but here we propose that it can be used to deliver very high power factors as well. Combining the two, could lead to exceptionally high *ZT*s.

Potential barriers can be optimized for achieving power factor improvements through energy filtering of the order of 30%. Nanoinclusions, on the other hand, have a weaker influence on the Seebeck coefficient and provide somewhat smaller improvements to the power factor of around 10%. Their utilization of energy filtering is not effective and cannot provide higher power factors compared to an optimized structure without nanoinclusions. Importantly, however, we have shown how the mild power factor improvements can be independent of the nanoinclusion density. In hierarchical nanostructured materials, the combination of nanoinclusions and superlattice-like barriers can be designed for the PF to be almost completely immune to the presence and density of nanoinclusions. These findings indicate that larger densities of nanoinclusions can be utilized to effectively reduce the lattice thermal conductivity without degradation in the power factor.

We have then shown that very high power factors can be achieved once the grain/grain-boundary (well/barrier) interface is properly optimized, and combined with ideas from modulation doping. Specifically, the combination of energy filtering from a properly-designed well/barrier region, with dopant non-uniformity providing dopant-clean regions in which carriers have close to phonon-limited mobility, is the key in largely improving the power factor. We have also shown that an essential ingredient is that the energy of the carriers does not relax significantly in the well regions in order to transfer the high Seebeck coefficient of the barriers into the wells. It is shown, however, that this is the most probable and realistic case if the well region sizes are in the order of a few tens of nanometers. Power factors beyond 30 mW/mK$^2$ can be reached, and we pointed out to some experimental evidence that supports this [76, 77, 79, 80].

Although some of the parameters the simulations employ in our theoretical exploration are relevant to Si, the design direction we propose can be widely applied to other materials as well. A large variety of materials and their alloys are undergoing nanostructuring through top-down approaches primarily to reduce their thermal conductivity. A few extra steps like setting the doping such that the Fermi level is at the barrier level and having grains of the order of a few optical mean-free-path distances, would provide decent power factor improvements. Bottom-up approaches can provide the flexibility to further allow more precise geometry formation and dopant placement. In any case, admittedly, precise control can be challenging in experiments, but not all of the 'ingredients' we present need to coexist to improve the power factor.

For the simulations and designs we have presented, we have used the fully quantum mechanical Non-Equilibrium Green's Function method, and calculated the thermoelectric power factor of 2D nanoribbon channels with embedded nanoinclusions modelled as potential barriers. This method is most relevant, as it captures all geometry details (with the 2D treatment limitation), important quantum mechanical effects such as tunneling and subband quantization, as well as relevant transport regimes from diffusive to ballistic, and coherent to incoherent. These are all important features that affect transport through such structures and need to be captured for an accurate understanding of their thermoelectric properties as we showed in the results throughout the paper. Thus, this method avoids approximations in geometry and in essential transport features. Similarly, we discussed briefly that Monte Carlo simulations can also prove very useful for optimizing electronic and thermoelectric transport in hierarchically nanostructured materials. In that case, although quantum effects are not captured, much larger channels with larger degree of disorder can be simulated.

The design guidelines we presented can prove useful in the design of nanostructured thermoelectric materials which provide not only low thermal conductivities, but high power factors as well, able to lead the leap forward for the thermoelectric materials technology.

# Acknowledgements

This work has received funding from the European Research Council (ERC) under the European Union's Horizon 2020 Research and Innovation Programme (Grant Agreement No. 678763).

Figure 1:

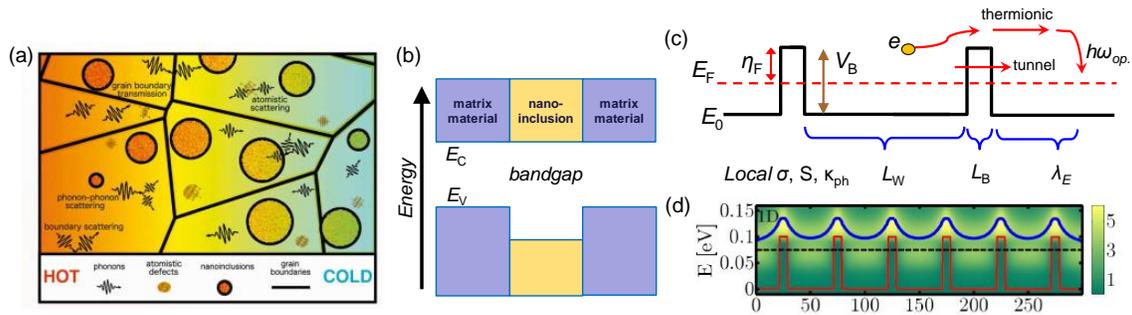

Figure 1 caption:

(a) Schematic of a hierarchically nanostructured material consisting of a matrix material with embedded atomic defects, nanoinclusions (NIs), and grain boundaries [32]. Phonons underdo scattering on all these inclusions. (b) The potential change in the vicinity of the NIs. Potential barriers can form, while band alignment is most beneficial for the electrical conductivity. (c) Illustration of charge transport in a superlattice structure, where electrons flow over potential barriers and relax into potential wells. (d) Quantum mechanical electronic transport simulation showing the energy of the current flow (blue line) in a superlattice material (red line). The dashed line indicates the Fermi level. Reprinted (d) from Ref. [181] with the permission of AIP Publishing.

Figure 2:

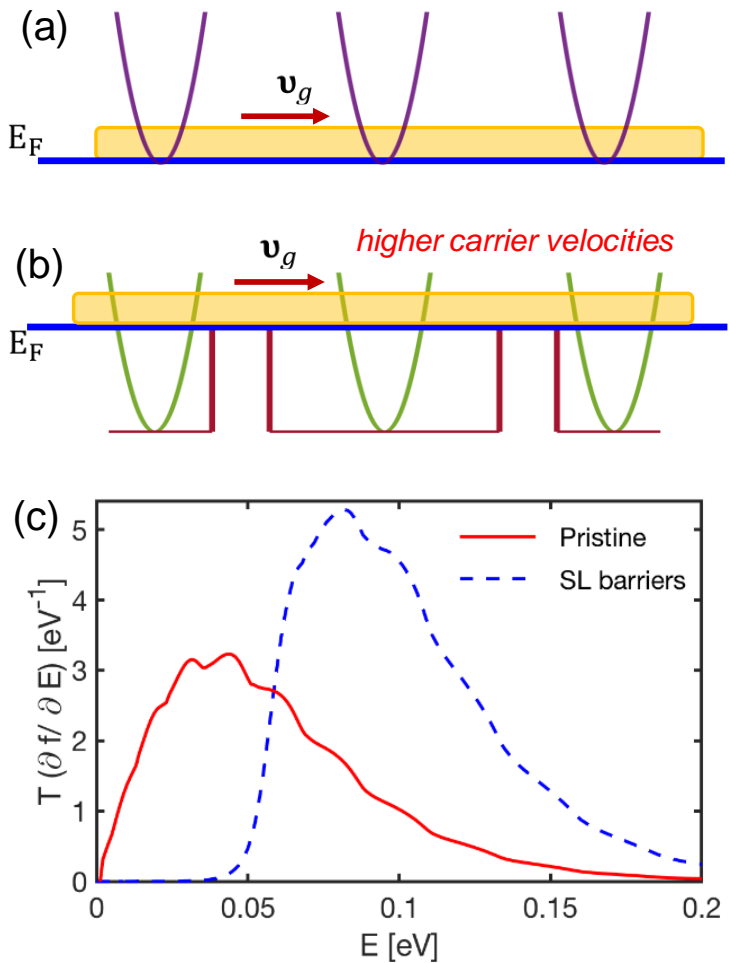

Figure 2 caption:

(a) Illustration of the current flow in a pristine material with the bandstructure of the material (parabolas), indicating the group velocity (slope of the parabolas). The Fermi level $E_F$ is aligned with the band edge. (b) Illustration of elastic current flow in a superlattice material (in the absence of energy relaxation), indicating that the group velocities of the electrons are larger compared to those of the pristine material as higher energy states are utilized for transport over the barrier. In this case the Fermi level $E_F$ is aligned with the barrier height. (c) The quantum mechanical transmission function of the electrons in the two cases, weighted by the derivative of the Fermi distribution, as appears in linearized transport and determines the current in the two materials. Reprinted (c) from reference [71] with permission from APS.

Figure 3:

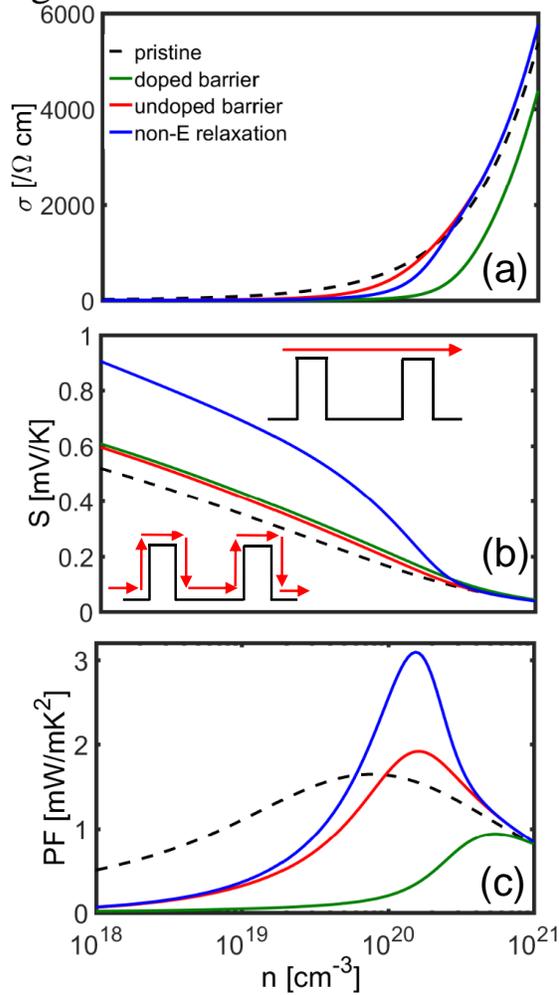

Figure 3 caption:

Thermoelectric coefficients for cases of materials and superlattice specifications (a) electrical conductivity, (b) Seebeck coefficient, (c) power factor versus carrier density. Parameters that reproduce the mobility of p-type Si are employed. Dashed-black lines: The pristine channel material without any barriers. Green lines: Superlattice channel for which the well/barrier regions are both doped. Red lines: Superlattice channel for which the barrier region is undoped and full energy relaxation is considered in the well region. Blue lines: The undoped barrier case as before, but with completely unrelaxed current energies in the wells. The barrier height is $V_B = 0.15$ eV. The schematics in (b) show the two current cases described by the red and blue lines.

Figure 4:

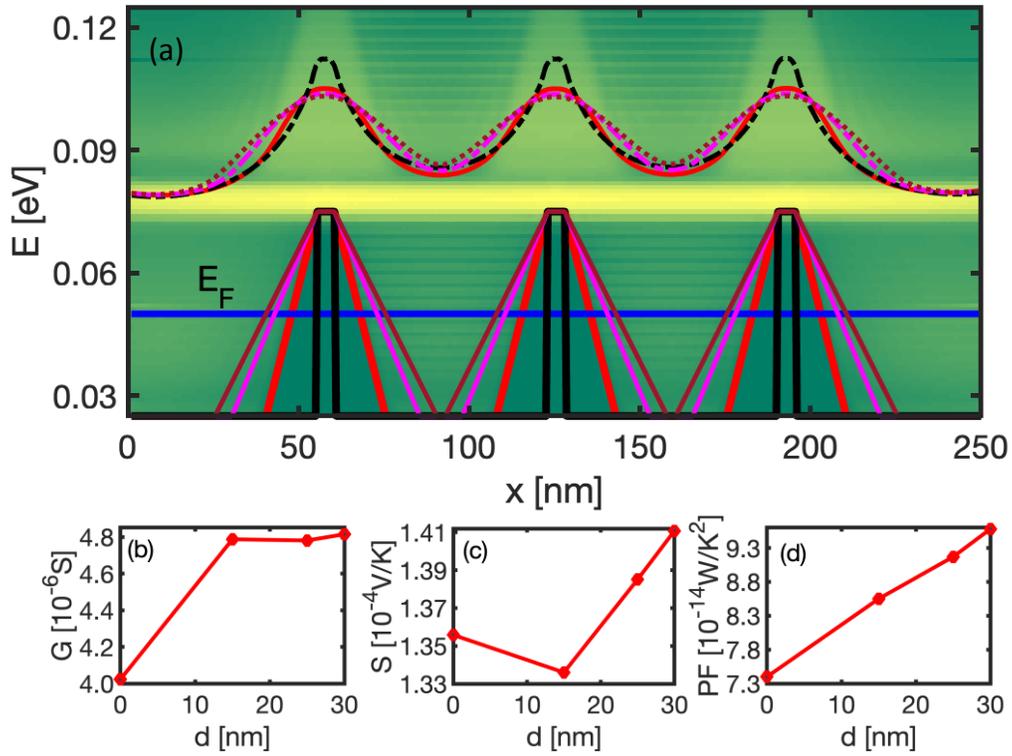

Figure 4 caption:

(a) The energy dependence of the current flow (yellow-green colormap) in superlattice (SL) structures with oblique barrier sidewalls, and the average energy of the current flow $<E(x)>$ (curved lines). The Fermi level is depicted by the flat blue line. The coloring of $<E>$ corresponds to the coloring of the barriers. (b-d) The conductance, Seebek coefficient, and power factor of the SLs as a function of the sidewall inclination distance, $d$. Adopted from Ref. [81].

Figure 5:

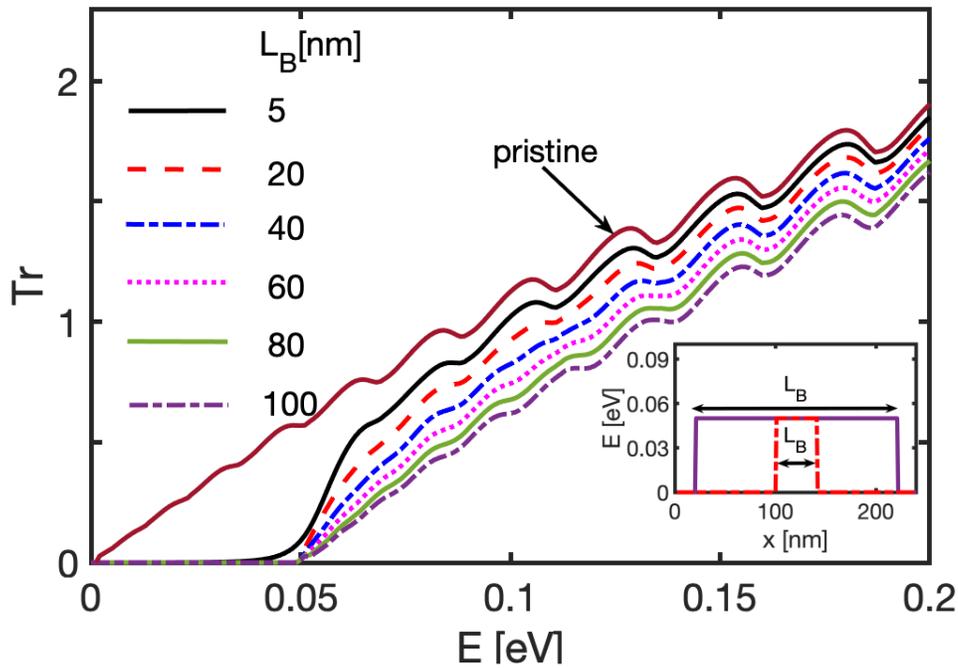

Figure 5 caption:

Indication of the degree of relaxation of carriers on the barrier material as they propagate over it. The figure shows the NEGF calculated energy resolved transmission function $Tr$ of the carriers in a channel with a single potential barrier with length $L_B$ as indicated in the inset. Electron scattering with acoustic phonons only are considered in the calculation. Cases for different barrier lengths are shown from a large $L_B = 100$ nm taking over most of the channel (purple-dashed line), to a pristine channel (brown line). The 'jump' of the $Tr$ from that of the larger barrier to that of the pristine material would indicate that carriers are more thermionically emitted over the barrier rather than 'relaxing on it' as the barrier length is scaled. Adopted from Ref. [81].

Figure 6:

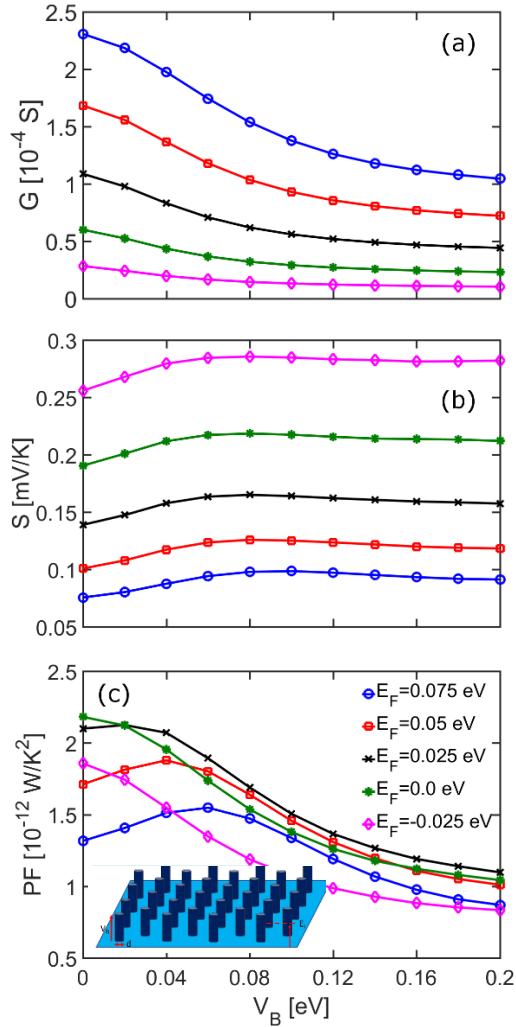

Figure 6 caption:

The thermoelectric coefficients of an $L = 60$ nm channel with an 8×4 hexagonal arrangement of nanoinclusions (inset of (c)) and acoustic phonon scattering transport conditions versus nanoinclusion barrier height, $V_B$. (a) The conductance. (b) The Seebeck coefficient. (c) The power factor defined as $GS^2$. Five different Fermi levels are considered: $E_F = -0.025$ eV (purple-diamond lines), $E_F = 0$ eV (green-star lines), $E_F = 0.025$ eV (black-cross lines), $E_F = 0.05$ eV (red-square lines), and $E_F = 0.075$ eV (blue-circle lines). Reprinted from reference [183] with permission from APS.

Figure 7:

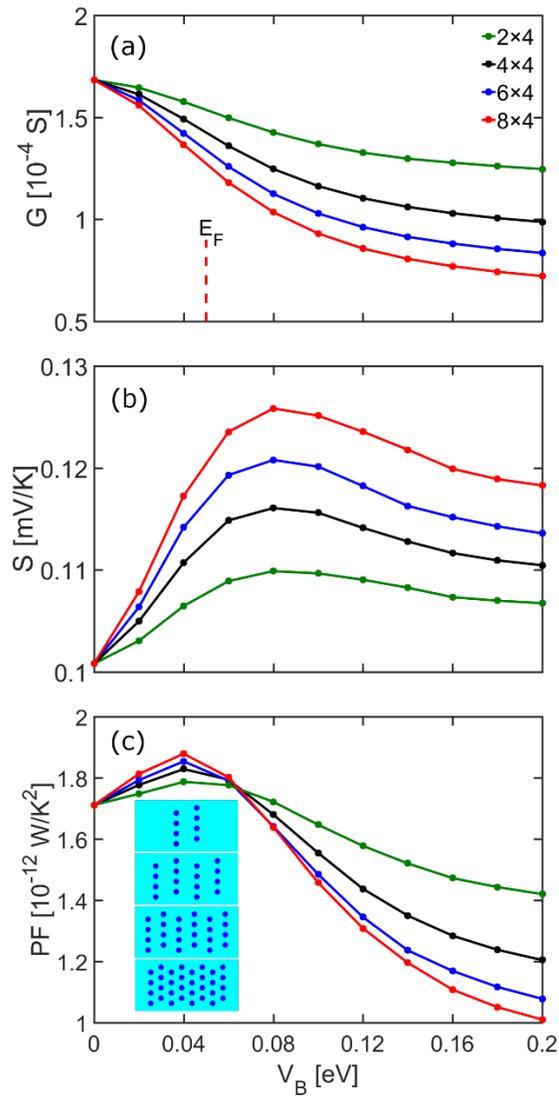

Figure 7 caption:

The thermoelectric coefficients of an $L$ = 60 nm channel with $E_F$ = 0.05 eV (dashed-red line) and acoustic phonon scattering transport conditions versus nanoinclusion barrier height, $V_B$. (a) The conductance. (b) The Seebeck coefficient. (c) The power factor defined as $GS^2$. Hexagonal arrays of four different nanoinclusion densities are considered as shown in the inset of (c): 2×4 array (green lines), 4×4 array (black lines), 6×4 array (blue lines), and 8×4 array (red lines). Reprinted from reference [183] with the permission of AIP Publishing.

Figure 8:

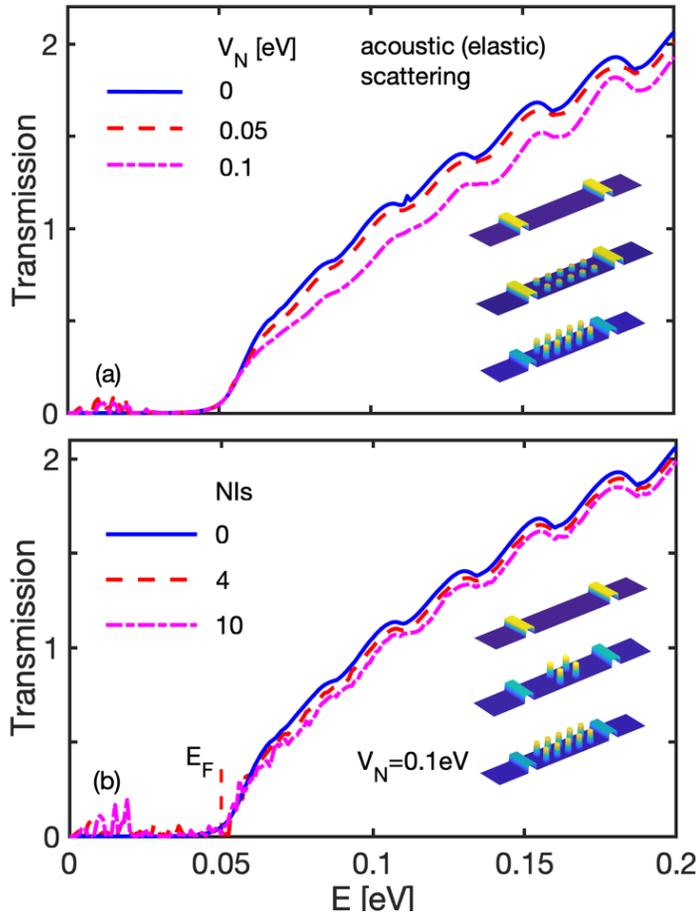

Figure 8 caption:

Transmission versus electron energy $E$ for a channel with SL barriers and nanoinclusions (NIs) in the elastic scattering regime (acoustic phonon scattering only) for: (a) increasing NI barrier height $V_N$, and (b) increasing number of NIs. In (b) the height of the NIs is set to VN = 0.1 eV. The insets show schematics of the channels considered. Reprinted from Ref. [71] with permission from APS.

Figure 9:

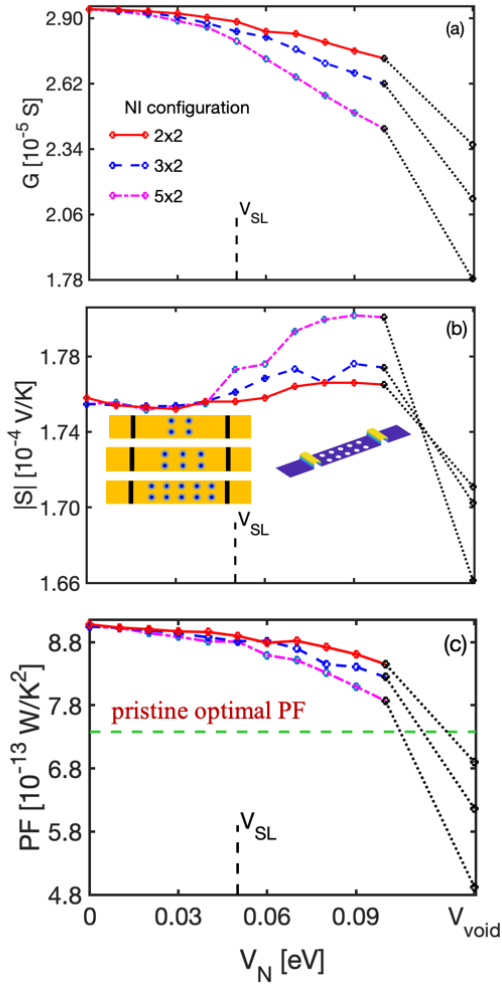

Figure 9 caption:

Conductance $G$, Seebeck coefficient $S$, and power factor PF, for SL structures with barrier nanoinclusions (NIs) as shown in the inset of (b). Results for increasing NI barrier heights $V_N$ are presented under electron-acoustic phonon limited conditions. In (c), the PF of the optimized pristine channel is shown by the green dashed line. The red, blue and magenta lines show results in which the number of NIs increases from 2 x 2 to 3 x 2 and 5 x 2 NIs, respectively-as shown in the inset of (b). The dotted black lines extend the results to the case where the NIs are replaced with voids, as shown in the inset of (b) as well. The SL barrier height, $V_{SL}$, is denoted as well. Reprinted from reference [71] with permission from APS.

Figure 10:

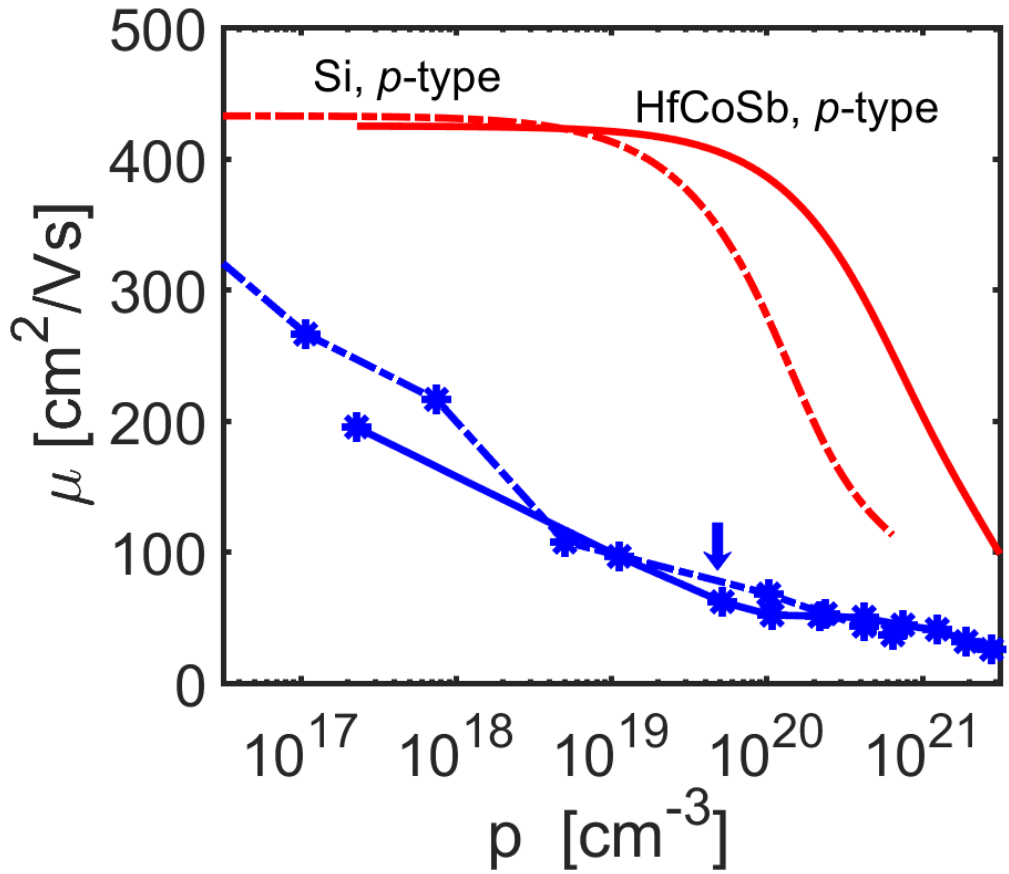

Figure 10 caption:

The mobility of p-type Si (dashed lines) and the HfCoSb half Heusler alloy (solid lines), which is a promising thermoelectric material. The red lines indicate the phonon-limited mobilities, whereas the blue lines the phonon plus ionized impurity scattering limited mobilities.

Figure 11:

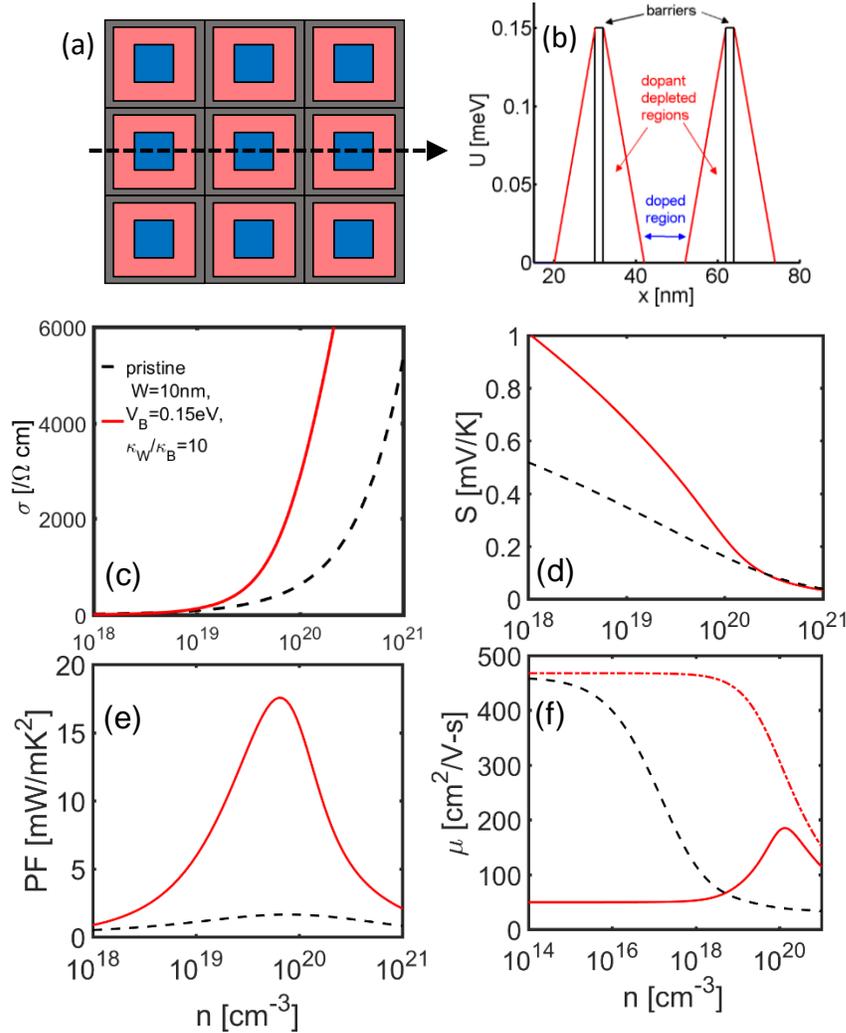

Figure 11 caption:

(a) A three-region structure design in a 2D top-down representation for high PFs. The core of the well region (blue regions) is highly doped, the intermediate region between the core and the boundary (red region) is intrinsic (dopant-free), and the region in-between the wells is shown in grey. (b) A 'cut' through the dashed line of (a), indicating a simplification of the conduction band profile, with the doped regions, the intrinsic regions and the boundary barrier regions indicated. (c-e) Thermoelectric coefficients (c) electrical conductivity, (d) Seebeck coefficient, (e) power factor versus carrier density for the pristine material (dashed-black lines) and the barrier material (solid-red lines). The barrier material is simulated for a depletion region width of W= 10 nm, a barrier height of $V_B = 0.15$ eV, and ratio of thermal conductivities between the well and barrier of 10. Parameters that reproduce the mobility of p-type Si are employed, whereas with dashed-red we show the phonon limited mobility of the pristine channel. (f) The mobility of the two channels. (a) and (b) are adopted from Ref. [81].

Figure 12:

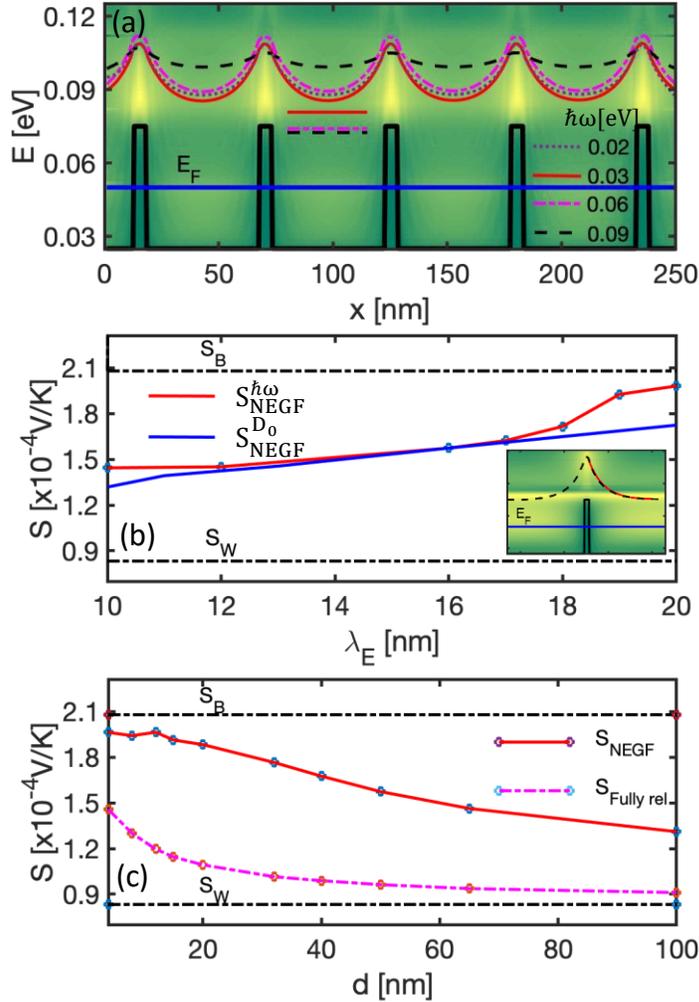

Figure 12 caption:

(a) The superlattice structure with the energy of the current flow <E> for different phonon energies 0.02 eV, 0.03 eV, 0.06 eV, and 0.09 eV. The short horizontal dashed lines indicate the energy of the current level in the pristine channel, i.e. where <E> would relax to in an infinitely long well. The colormap indicates the current flow $I(E,x)$. (b) The Seebeck coefficient versus the energy relaxation length of a superlattice for the case where $\lambda_E$ is altered by changing the phonon-energies (red line), and by altering the electron-phonon coupling strength (blue line). Inset: The extraction of $\lambda_E$ by an exponential fit of <E(x)> after the current passes over a single barrier. (c) The Seebeck coefficient of the superlattice structure versus the length of the well $d$ (red line). The magenta line shows the Seebeck coefficient in the case of full and immediate relaxation after the carriers pass over the barriers for the case of 0.06 eV phonon energy. The dashed horizontal lines in (b) and (c) indicate the Seebeck coefficient of a pristine material without barriers $S_W$ (infinite well), and of a pristine material with a large barrier $S_B$ (infinite barrier). Reprinted from reference [182] with the permission of AIP Publishing.

Figure 13:

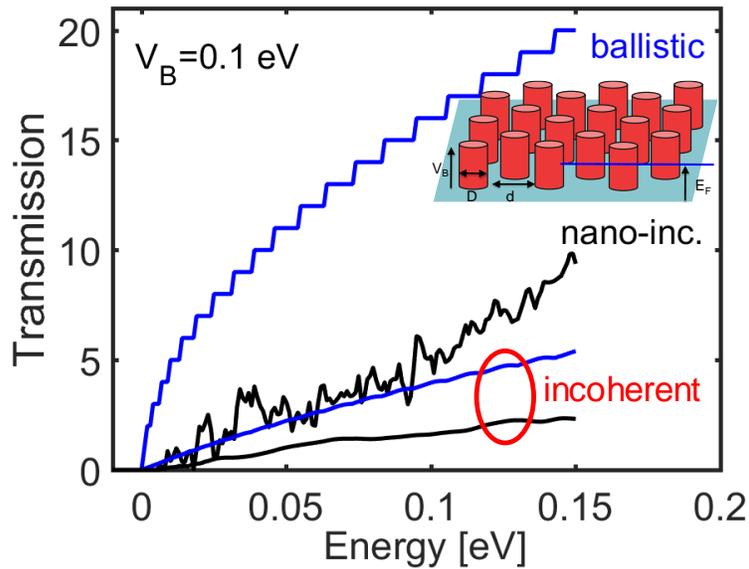

## Figure 13 caption:

The transmission versus energy for an $L = 60$ nm channel with nanoinclusions (as shown in the inset) for four geometry and transport cases: Blue-staircase line: Pristine channel under ballistic coherent (no phonon scattering) conditions. Blue-almost linear line: Pristine channel under diffusive (electron-phonon scattering) conditions. Black-edged line: Channel with nanoinclusions under coherent (no phonon scattering) conditions. Black-almost linear line: Channel with nanoinclusions under diffusive (electron-phonon scattering) conditions.